\documentclass[twocolumn]{aastex63}

\usepackage{amsmath}
\hypersetup{linkcolor=red,citecolor=magenta,filecolor=cyan,urlcolor=magenta}

\submitjournal{AJ}

\shorttitle{Planet Occurrence Rate Correlated to Stellar Dynamical History}
\shortauthors{Yuan-Zhe Dai et al.}
\graphicspath{{./}{figures/}}

\begin{document}

\title{Planet Occurrence Rate Correlated to Stellar Dynamical History: Evidence from Kepler and Gaia}

\correspondingauthor{Hui-Gen Liu}
\email{huigen@nju.edu.cn}

\author{Yuan-Zhe Dai}
\affiliation{School of Astronomy and Space Science, Nanjing University, 163 Xianlin Avenue, Nanjing, 210023, People's Republic of China}
\affiliation{Key Laboratory of Modern Astronomy and Astrophysics, Ministry of Education, Nanjing, 210023, People's Republic of China}
\author{Hui-Gen Liu}
\affiliation{School of Astronomy and Space Science, Nanjing University, 163 Xianlin Avenue, Nanjing, 210023, People's Republic of China}
\affiliation{Key Laboratory of Modern Astronomy and Astrophysics, Ministry of Education, Nanjing, 210023, People's Republic of China}
\author{Dong-Sheng An}
\affiliation{School of Astronomy and Space Science, Nanjing University, 163 Xianlin Avenue, Nanjing, 210023, People's Republic of China}
\affiliation{Key Laboratory of Modern Astronomy and Astrophysics, Ministry of Education, Nanjing, 210023, People's Republic of China}
\author{Ji-Lin Zhou}
\affiliation{School of Astronomy and Space Science, Nanjing University, 163 Xianlin Avenue, Nanjing, 210023, People's Republic of China}
\affiliation{Key Laboratory of Modern Astronomy and Astrophysics, Ministry of Education, Nanjing, 210023, People's Republic of China}





\begin{abstract}
The dynamical history of stars influences the formation and evolution of planets significantly. To explore the influence of dynamical history on the planet formation and evolution from observations, we assume stars that experienced significantly different dynamical histories tend to have different relative velocities. Utilizing the accurate Gaia-Kepler Stellar Properties Catalog, we select single main-sequence stars and divide these stars into three groups according to their relative velocities, i.e. high-\textit{V} , medium-\textit{V} , and low-\textit{V} stars. After considering the known biases from Kepler data and adopting prior and posterior correction to minimize the influence of stellar properties on planet occurrence rate, we find that high-\textit{V} stars have a lower occurrence rate of super-Earths and sub-Neptunes (1--4 $R_{\oplus}$, P$\textless$100 days) and higher occurrence rate of sub-Earth (0.5--1 $R_{\oplus}$, P$\textless$30 days) than low-\textit{V} stars. Additionally, high-\textit{V} stars have a lower occurrence rate of hot Jupiter sized planets (4--20 $R_{\oplus}$, P$\textless$10 days) and a slightly higher occurrence rate of warm or cold Jupiter sized planets (4--20 $R_{\oplus}$, 10$\textless$P$\textless$400 days). After investigating the multiplicity and eccentricity, we find that high-\textit{V} planet hosts prefer a higher fraction of multi-planets systems and lower average eccentricity, which is consistent with the eccentricity-multiplicity dichotomy of Kepler planetary systems. All these statistical results favor the scenario that the high-\textit{V} stars with large relative velocity may experience fewer gravitational events, while the low-\textit{V} stars may be influenced by stellar clustering significantly.
\end{abstract}



\section{Introduction} \label{intro}
Since the first discovery of an exoplanet around a solar-type star in 1995 \citep{1995Natur.378..355M}, more than 4300\footnote{https://exoplanetarchive.ipac.caltech.edu/} exoplanets detected. Kepler space telescope has discovered more than 2291 exoplanets and 1786 candidates (based on Kepler DR25 \cite{2018ApJS..235...38T}), a rich transiting exoplanet sample via single telescope. To refine the parameters of Kepler planets, different groups have done spectral follow-ups, e.g. California-$Kepler$ Survey (CKS) \citep{2017AJ....154..108J} and the Large Sky Area Multi-Object Fiber Spectroscopic Telescope survey (LAMOST) \citep{2012RAA....12.1197C,2012RAA....12..723Z,2012RAA....12.1243L,2015RAA....15.1095L}. Thanks to the precise astrometric data from Gaia DR2 \citep{2018A&A...616A...1G}, the stellar parameters can be refined more accurately, e.g. effective temperature and stellar radius. Consequently, the more accurate the planet parameters will be. With these accurate planet parameters, we are now in a great epoch to explore the correlations between stellar parameters and planet parameters or planetary system architectures. 

Stars usually form in clustering environments \citep{1993prpl.conf..245L,2000AJ....120.3139C,2003ARA&A..41...57L}. Due to the galactic tides, most clustering stars would become field stars eventually. The differences between the occurrence rate of planets around stars in clusters and field stars are crucial to understanding the planetary formation and evolution in cluster environments. From the aspect of observation, several surveys have monitored stars in young, metal-rich open clusters and old, metal-poor globular clusters. However, only tens of exoplanets have been found in open clusters, and two pulsars as planet hosts are detected in globular clusters(GCs). Simply comparing the total number of planets in clusters and field stars(considering most of Kepler stars are field stars), planets in clusters are only a small fraction of the known exoplanet sample. It seems that planets are rare in clusters. Some recent works hold a different view. \cite{2011ApJ...729...63V} indicated that low detection probabilities of planets in distant open clusters may cause a significant lack of hot Jupiters, compared with hot Jupiters around field stars. \cite{2013Natur.499...55M} suggested that both the orbital properties and the frequency of planets in open clusters are consistent with those in the field of the Milky Way. \cite{2017A&A...603A..85B} found that the occurrence of the planet rate of hot Jupiters in open clusters is slightly higher than that in the field. All these previous planet searches in open clusters indicated that planets discovered in open clusters appear to have properties in common with those found around the field stars. 

Theoretically, both the UV radiation from nearby stars \citep{1998ApJ...499..758J,2003ApJ...582..893M,2018MNRAS.480.4080D,2018MNRAS.478.2700W,2020A&A...640A..27V} and gravitational perturbation during the frequent close stellar encounters \citep{2006ApJ...642.1140O,2009ApJ...697..458S,2013ApJ...772..142L,2017MNRAS.470.4337C,2017AJ....154..272H} will probably affect the planets formation and evolution considering the complicated clustering environment. In the planet-forming disks, disk dispersal is essential, especially for the formation of gas giants. According to the classical core-accretion model \citep{2004ApJ...616..567I}, the disk lifetime will determine whether a proto-planet can grow up to a gas giant or how massive the planet can grow finally. We proposed a viscous photo-evaporative disk model combining the photo-evaporation of external flyby stars and host stars \citep{2018MNRAS.480.4080D}. Additionally, we applied this model to the clustering environment to explain rare gas giants in dense globular clusters or clusters with massive stars. Apart from the radiation environments in clusters, the gravitational perturbations of external stars also affect planet formation both in the stages of gas disk evolution and afterward. \cite{2006ApJ...642.1140O} showed that nearly 90\% of the disk mass will be removed during a close encounter in extreme cases. \cite{2009ApJ...697..458S,2013ApJ...772..142L}, use N-body simulation shows that the instability of planets in clusters is considered to be an important role. A large fraction of planets very close to the host star are probably stable in open clusters or even in the outer region of GCs. Especially, \cite{2019A&A...624A.110F} showed that the planet survival rate of planets in clusters would decrease with increasing semi-major axis.

Based on current studies, it is under debate that the occurrence rate of planets around stars in open clusters and around field stars are the same or not. Due to the small number of planets in stellar clusters, the subsequent large statistical uncertainties can hardly achieve conclusive results. However, Gaia provides unique data to describe the stellar motions due to the extremely high-precision astrometric data. Utilizing Gaia DR2 \citep{2018A&A...616A...1G}, we can achieve the accurate stellar motion parameters in the Kepler field. 

Gaia DR2 provides accurate astrometric data. Kepler DR25 provides planets properties. Utilizing Gaia DR2 and planet host stars, \cite{2020Natur.586..528W} finds that stellar clustering shapes the architecture of planetary systems. \cite{2020ApJ...905L..18K} argues that stellar clustering is a key driver for turning sub-Neptunes into Super-Earths. These two works provide us a new window connecting planet formation and evolution, star and stellar cluster formation, and galaxy evolution. Similarly, using Gaia DR2 and Kepler data, we can study the correlations between stellar relative velocity and planets occurrence rate(the average number of planets per star).         
 
Recently, two groups investigated the correlations between stellar motion and planet occurrence. \citet{2019AJ....158...61B} used a cross-matched catalog which includes Gaia DR2, Kepler DR25, and LAMOST DR4\footnote{http://dr4.lamost.org/}. They found that planets around stars with lower iron metallicity and higher total speed tend to have a higher occurrence rate. However, due to the limited number of stars with three-dimensional velocities, these correlations may be coupled with the influence of effective temperature and other properties i.e. they may be biased. 

\citet{2019MNRAS.489.2505M} also investigates the correlation between velocity of Kepler field stars and Kepler planet hosts and argued that planet occurrence is independent with stellar velocity. Actually they haven't calculated the occurrence rate of planets around stars with difference velocity. Besides that they haven't considered the potential observational biases from the Kepler data, thus, it's necessary to revisit the correlations between stellar velocity and planet occurrence rate. 

Additionally, there are some other factors that need to be included, e.g. the effective temperature and metallicity of stars. \cite{2012ApJS..201...15H,2015ApJ...798..112M} found that the occurrence rate of planets with radius of 1--4 $R_{\oplus}$ will decrease with the increasing of effective temperature of stars. In other words, planets orbit more common around cool stars. Several works all show that occurrence rate of planet has a positive correlation with stellar metallicity \citep{2004ApJ...616..567I,2015AJ....149...14W,2016AJ....152..187M,2019ApJ...873....8Z} which supports the classical core-accretion scenario. Here if we use the proper motions and parallax of all the Kepler stars, we can get two-dimensional projected velocities in barycentric right ascension (RA) and barycentric declination(DEC). Comparing these two-dimentional velocities with nearby stars, we can get their relative velocities and make statistical analysis based on their relative velocities. Due to previous studies on correlations between stellar properties and planet occurrence rate, we need to deal with the influence of stellar properties very carefully when investigating the correlation between planet occurrence rate and stellar relative velocity.

This paper will be organized as follows, in section \ref{methodology}, we introduce our methodology, i.e. in subsection \ref{sub 2.1} we introduce selection of main-sequence stars and planets around them; in subsection \ref{sub 2.2}, we describe the calculation of two-dimensional velocity and split the stellar samples according to their relative velocities; In section \ref{results of focc}, we show our main results, i.e. the correlations between stellar properties and relative velocities; the correlations between occurrence rate of different sized planets and stellar relative velocities; the correlations between eccentricity, multiplicity, and stellar relative velocities. In section \ref{scenarios}, we discuss several scenarios to explain our statistical results. The conclusions are summarized in section \ref{conclusion}.

\section{Methodology} \label{methodology}
\subsection{sample selection} \label{sub 2.1} 
The preliminary catalog we used is based on the table from \cite{2020AJ....159..280B}, the Gaia-Kepler Stellar Properties Catalog, which is a set of the most accurate stellar properties, and homogeneously derived from isochrones and broadband photometry. Since we aim to unearth the correlations between stellar relative velocity and planet occurrence rate, we need to obtain the two-dimensional velocity of Kepler stars. The criteria of stellar samples selected to calculate two two-dimensional velocities are as follows: 
\begin{itemize}
    \item 1. The selected stars are probably main-sequence stars.
    \item 2. The selected stars are probably single stars.
\end{itemize}
We choose main-sequence stars to exclude the potential systematic biases depending on the stellar evolving stage, whereas we choose single stars because the gravitational effects of stars in binary systems or multi-stars systems may influence the stellar proper motion. Before the calculation of relative velocities, we define nearby stars, i.e. stars around a chosen Kepler star within 100 pc. Then we derive the average velocity and velocity dispersion of these stars around a given Kepler target.

\cite{2018ApJ...866...99B}(hereafter B2018) revised the stellar radii using the astrometry and photometry data from Gaia DR2. Here, we exclude stars flagged as sub-giants or red giants in \cite{2018ApJ...866...99B} and only choose main-sequence stars. We also exclude the potential cool binary stars due to their inflated radii. Additionally, according to \cite{2020AJ....159..280B}(hereafter B2020), we remove some Kepler stars, around which Gaia-detected companions within 4 arcsec may be binaries that will contaminate secondary K$_{\rm s}$ magnitudes. What's more, \cite{2020AJ....159..280B} suggests that stars with RUWE\footnote{RUWE is the magnitude- and color independent re-normalization of the astrometric $\chi^{2}$ of Gaia DR2 (unit-weight error of UWE)} $\gtrsim$ 1.2, which are likely to be binaries(A. Kraus et al., in prep). To be cautious with the sample selection, we exclude all of these stars with RUWE $\gtrsim$ 1.2. Because binary stars may not only influence our calculation of velocity dispersion but also intrinsically have significant impacts on planet formation. For instance, planets around binary stars are different from single stars. E.g. the tidal truncation of the gas disk by companion stars \citep{2010ApJ...708.1566X,2015ApJ...798...71S}, meanwhile, the gravitational secular perturbation may significantly change the architectures of planetary systems, e.g. Kozai-Lidov oscillations coupled to tidal friction for close binaries \citep{2016ApJ...827....8N,2019MNRAS.485.4967F}.  

To sum up, we choose three criteria for excluding potential binaries or stars in multi-star systems i.e. 1. Stars flagged as cool binaries in B2018; 2. Stars flagged with RUWE $\gtrsim$ 1.2 in B2020; 3. Stars with contaminated K$_{\rm s}$ mag.


\begin{figure*}
    \centering
    \includegraphics[width=1\linewidth]{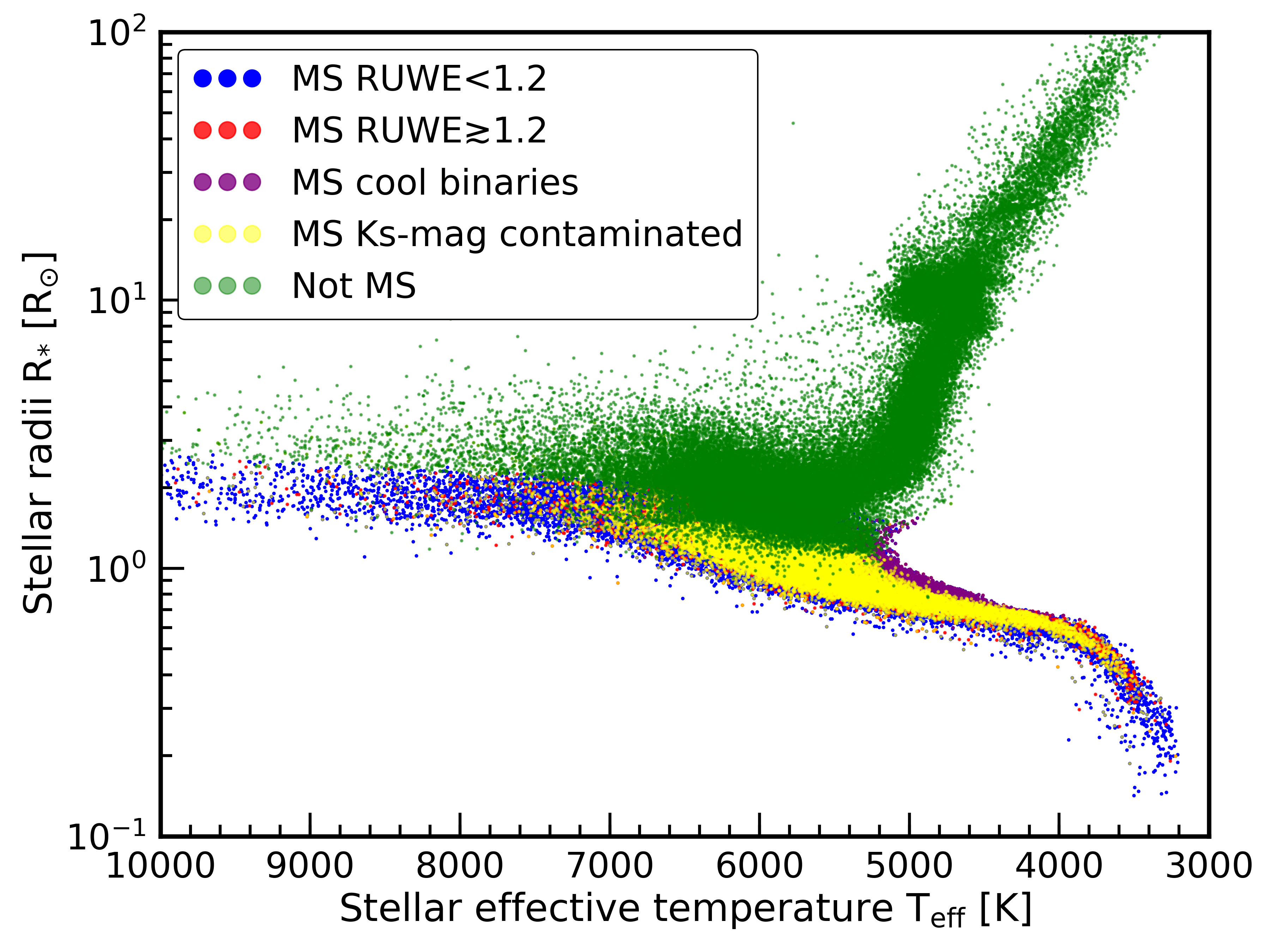}
    \caption{Evolutionary state classifications of all Kepler targets based on flags described in \cite{2018ApJ...866...99B} and \cite{2020AJ....159..280B}. We find that $\simeq$ 67.5\% of Kepler targets(stars in Kepler Input Catalogue) are main-sequence stars. $\simeq$ 62.0\% of Kepler targets are main-sequence stars with RUWE $\textless$ 1.2(blue), $\simeq$ 5.4\% of Kepler targets are main-sequence stars with RUWE $\gtrsim$ 1.2(red) and $\simeq$ 2\%(purple) of Kepler targets are main-sequence stars flagged as cool binary stars. Other evolved stars including subgiants and red giants make up $\simeq$ 32.5\%(green) of Kepler targets. We also select Kepler stars(yellow), around which Gaia-detected companions within 4 arcsec may be binaries and contaminate secondary Ks magnitudes.}
	\label{figure 2-1}
\end{figure*}

Fig \ref{figure 2-1} shows evolutionary state classifications of all Kepler targets based on flags described in \cite{2018ApJ...866...99B} and \cite{2020AJ....159..280B}. Here we select probably main-sequence stars according to B2018. The aim of B2020 is to rederive the stellar properties utilizing Gaia DR2 parallaxes, homogeneous stellar g and Ks photometry, and spectroscopic metallicity. B2020 provides evolved stars in the RGB stage or clump stage, however, it doesn't provide main-sequence stars. We use the main-sequence stars flagged in B2018. While we use B2020 to exclude stars that are likely in multi-stellar systems according to RUWE and whether the $K_{\rm s}$ mags of stars are contaminated. Although the binary main sequence stars identified in B2018 are not prominent in B2020, we cautiously exclude these cool binaries listed in B2018. In all stellar targets in Kepler Input Catalogue(KIC), we find that $\simeq$ 67.5\% of them are main-sequence stars. $\simeq$ 62.0\% of them are main-sequence stars with RUWE $\textless$ 1.2(blue), $\simeq$ 5.4\% of them are main-sequence stars with RUWE $\gtrsim$ 1.2(red) and $\simeq$ 2\%(purple) of them are cool binary main-sequence stars. Other evolved stars such as subgiants and red giants make up $\simeq$ 32.5\%(green) of Kepler targets. 
\begin{table}[]
\centering
\footnotesize
\caption{Sample selection with data of B2020}\label{table 1}

\begin{tabular}{c|cc}
\hline
                                             & Star    & Planet \\ \hline
Kepler DR25                                & 199991 & 8054   \\
Cross-matching with B2020                        & 186301 & ...    \\
Main sequence                                 & 119779 & ...    \\
Excluding cool binaries listed in B2018                                  & 116387 & ...    \\
Duty cycle\footnote{Duty cycle is the fraction of data cadences within the span of observations that contain test data and contribute toward detection of transit signals.}$\geq$ 0.6 and data span\footnote{The time elapsed in days between the first and last cadences containing test data.}\textgreater 2yr                            & 100594 & ....   \\
RUWE \textless 1.2 & 92681 & 4197 \\ 
Excluding stars with contaminated K$_{\rm s}$-mag \footnote{K$_{\rm s}$-mag flag such as "BinaryCorr" and "TertiaryCorr" indicates potential
binarity because that Gaia-detected companions within 4" for a given Kepler star may contaminate secondary Ks magnitudes.}& 77508 & 3579 \\ 
Distance $\leq$ 2000 pc & 75778 & 3533\\
FGK stars: 3850K $\leq T_{\rm eff}\leq$ 7500K & 74002 & 3441 \\ 
Deposition score \footnote{Deposition sore indicates the confidence in the KOI disposition.} $\geq$ 0.9                           & ...     & 1910   \\
\hline
\end{tabular}


\end{table}

Table \ref{table 1} lists the numbers of stars and planets after every selection step. We follow several steps -  cross-matching with B2020 (186301 stars left), selecting main sequence stars (119779 stars left), excluding cool binaries flagged in B2018 (116387 stars left), excluding stars with duty cycle $\geqslant$ 0.6 and data span $\textless$ 2 yr (100594 stars left) \citep{2018AJ....156..221N,2020AJ....159..164Y}, excluding stars with RUWE $\gtrsim$ 1.2 (92681 stars left) and excluding the stars with contaminated K$_{\rm s}$-mag flagged (77508 stars left). Around 77508 probably main-sequence single stars, there is 3579 corresponding Kepler Object of Interests(KOI) left. Furthermore, we exclude stars with distance larger than 2 kpc and select FGK stars (3850 K - 7500 K)  using the spectral–temperature relationship as in \cite{2013ApJS..208....9P} (74002 stars left). Here, we follow the criterion in \cite{2018AJ....156...24M} and select 1910 reliable planet candidates whose $Robovetter$ disposition scores are larger than 0.9.

\subsection{Stellar relative velocity} \label{sub 2.2}

\subsubsection{the calculation of velocity dispersion and relative velocity} \label{subsub 2.2.1}
Gaia DR2 provides us the most accurate astrometric data of Milky stars up to now. With these precise astrometric data including the position, distance, and proper motion of stars, we can easily calculate the velocity both in the right ascension direction and declination direction, i.e. $\upsilon_{\rm ra}$ and $\upsilon_{\rm de}$. 

\begin{equation} \label{equation 2-1}
    \upsilon_{\rm i} = \frac{\mu_{\rm i}}{\pi}
\end{equation}
where $i=ra, dec$ represents the direction of stellar velocity, i.e. right ascension direction and declination direction, respectively. $\mu$ is the proper motion, and $\pi$ is parallax.

Because we aim to explore the correlations between planet occurrence rate and stellar relative velocity, we need to calculate the stellar velocity relative to nearby stars. However, stars in the Milky Way have different rotational speeds with different distances according to the well-known Milky Way rotation curve \cite{2009PASJ...61..227S}. For most of the Kepler targets in the Kepler region, their distance from the center of the Milky Way is about 8--10 kpc. The difference of rotational speed may be up to several tens of kilometers per second (see the figure of Milky Way rotation speed curve in \cite{2009PASJ...61..227S}). So if we use all the stars in the Kepler region, some systematic bias will be brought into our calculation. Here when we calculate the relative velocity of a given Kepler star, we select its nearby stars within 100 pc to calculate the average velocity and the velocity dispersion of these stars. The formula of velocity dispersion is written as follows:

\begin{equation} \label{equation 2-2}
    \sigma_{\rm i} \equiv  \langle ( \upsilon_{\rm i} - \langle \upsilon_{\rm i} \rangle )^{2} \rangle^{1/2},
\end{equation}
where $\sigma_{\rm i}$ is the stellar velocity dispersion, i.e. the subscript $i=ra,dec$ represents the direction of right ascension(RA) and declination(DEC), respectively. $\upsilon_{\rm i}$ is the velocity in the direction of RA and DEC. Calculation of the velocity dispersion of nearby stars will minimize the effect of stellar rotation speed varying with the position in the Milky Way.

\subsubsection{The sample classification via relative velocity}
\label{subsub 2.2.2}
\begin{figure*}
    \centering
    \includegraphics[width=1\linewidth]{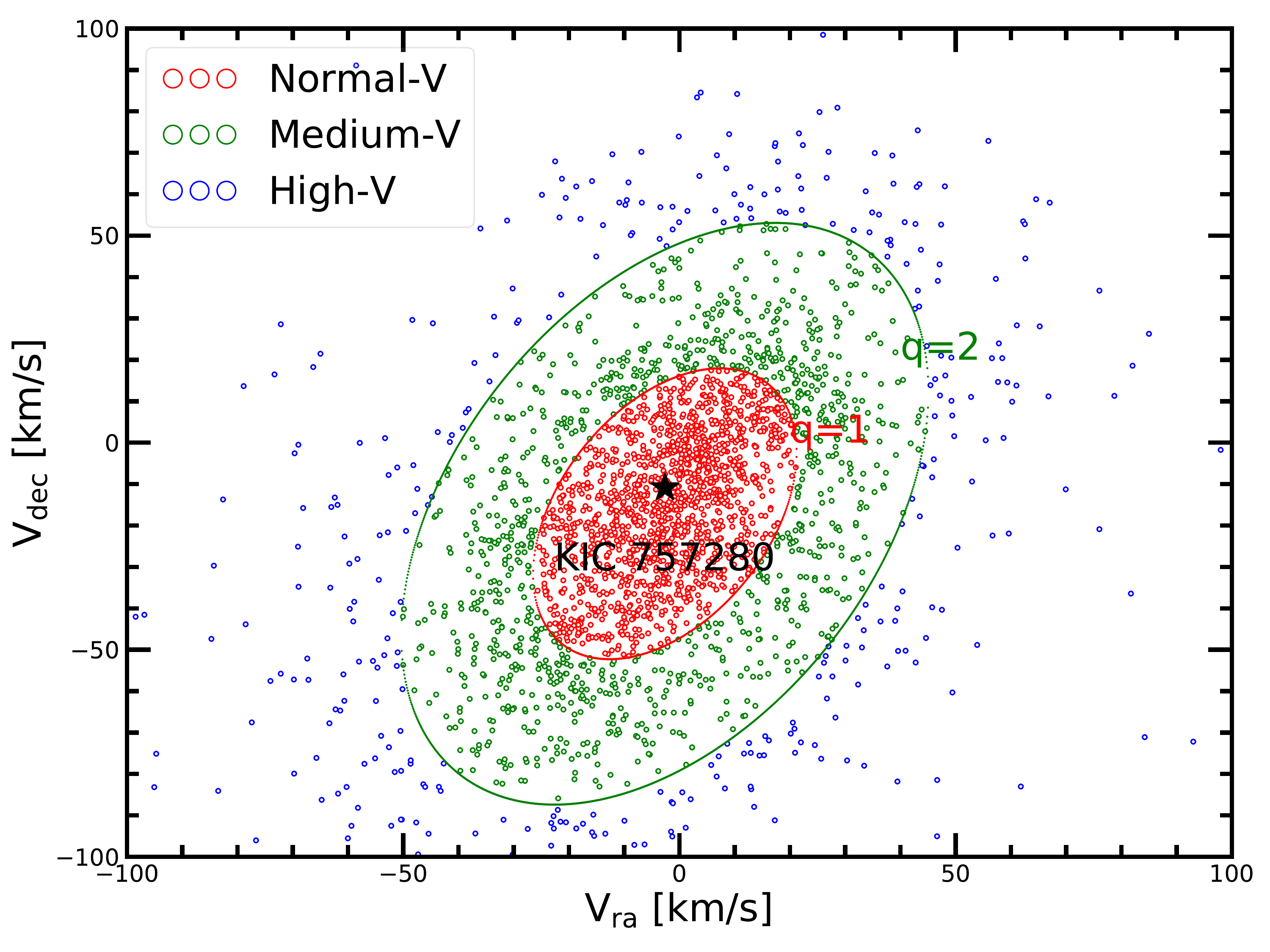}
    \caption{Take KIC 757280 as an example to show how we classify stars into three different groups according to relative velocities. We select stars within 100 pc of KIC 757280 from our catalog. KIC 757280 is indicated with a black star. Here high-\textit{V} stars(blue circles) are outside the green ellipse; medium-\textit{V} stars(green circles) are located between the red and green ellipse; low-\textit{V} stars(red circles) are inside the red ellipse.}
    \label{figure 2-3}
\end{figure*}

Here we define a new quantity $q$ that describes the deviation of stellar velocity relative to the average stellar velocity calculated in the range of stars around a given Kepler star. 

\begin{equation}
    \label{equation 2-3}
    q \equiv \sqrt{\left(\boldsymbol{V}-\boldsymbol{\mu}\right)^{T}\boldsymbol{C}^{-1}\left(\boldsymbol{V}-\boldsymbol{\mu}\right)}
\end{equation}

\begin{equation}
    \boldsymbol{V}=\left(\begin{matrix}V_{\rm ra} \\V_{\rm dec} \end{matrix}\right),\,\,
    \boldsymbol{\mu}=\left(\begin{matrix}\mu_{\rm ra} \\\mu_{\rm dec} \end{matrix}\right)
\end{equation}
where $\boldsymbol{V}$ is the 2D velocity vector, $\boldsymbol{\mu}$ is the average 2D velocity vector, and $\boldsymbol{C}$ is the covariance matrix of $\boldsymbol{V}$. 
We divide 74,002 single main-sequence stars from our catalog into three groups: high-\textit{V} stars (9754) i.e. stars with high relative velocity compared with stars in the proximity, medium-\textit{V} stars (28049) i.e. stars with median relative velocity, low-\textit{V} stars (36199) i.e. stars with low relative velocity. As is shown in Fig \ref{figure 2-3}, KIC 757280 is a low-\textit{V} star whose $0 \textless q \leqslant 1$.

\begin{itemize}
    \item high-\textit{V} stars: $q>2$;
    \item medium-\textit{V} stars: $1 \textless q \leqslant 2$;
    \item low-\textit{V} stars: $0 \textless q \leqslant 1$.
\end{itemize}

Additionally, previous researches on the solar system neighborhood have found that stellar velocity dispersion has correlations with stellar spectra type or effective temperature. More specifically, when the color index of the B-V mag is smaller than 0.61, stellar velocity dispersion has a strongly positive correlation with B-V mag; when the color index of B-V mag is large than 0.61, based on the data of Hipparcos, stellar velocity dispersion reaches a plateau stage \citep{1998MNRAS.298..387D}. Thus, considering these factors that may influence the calculation of velocity dispersion, we should have selected the nearby stars with similar effective temperature(deviation of $T_{\rm eff}$ is less than 500 K) within 100 pc around a given target star to estimate average velocity and dispersion. However, we do not add this criterion in our calculation. On the one hand, this selection criterion of a similar temperature will reduce the number of stars when calculating velocity dispersion, which will increase the statistical uncertainty. On the other hand, the majority($\sim$ 90\% ) of high-\textit{V} stars are the same, no matter we choose stars with or without the temperature criterion. Therefore, the different definitions of velocity dispersion only have limited influence on the identification of high-\textit{V} stars. 

\begin{figure*}
    \centering
    \includegraphics[width=1\linewidth]{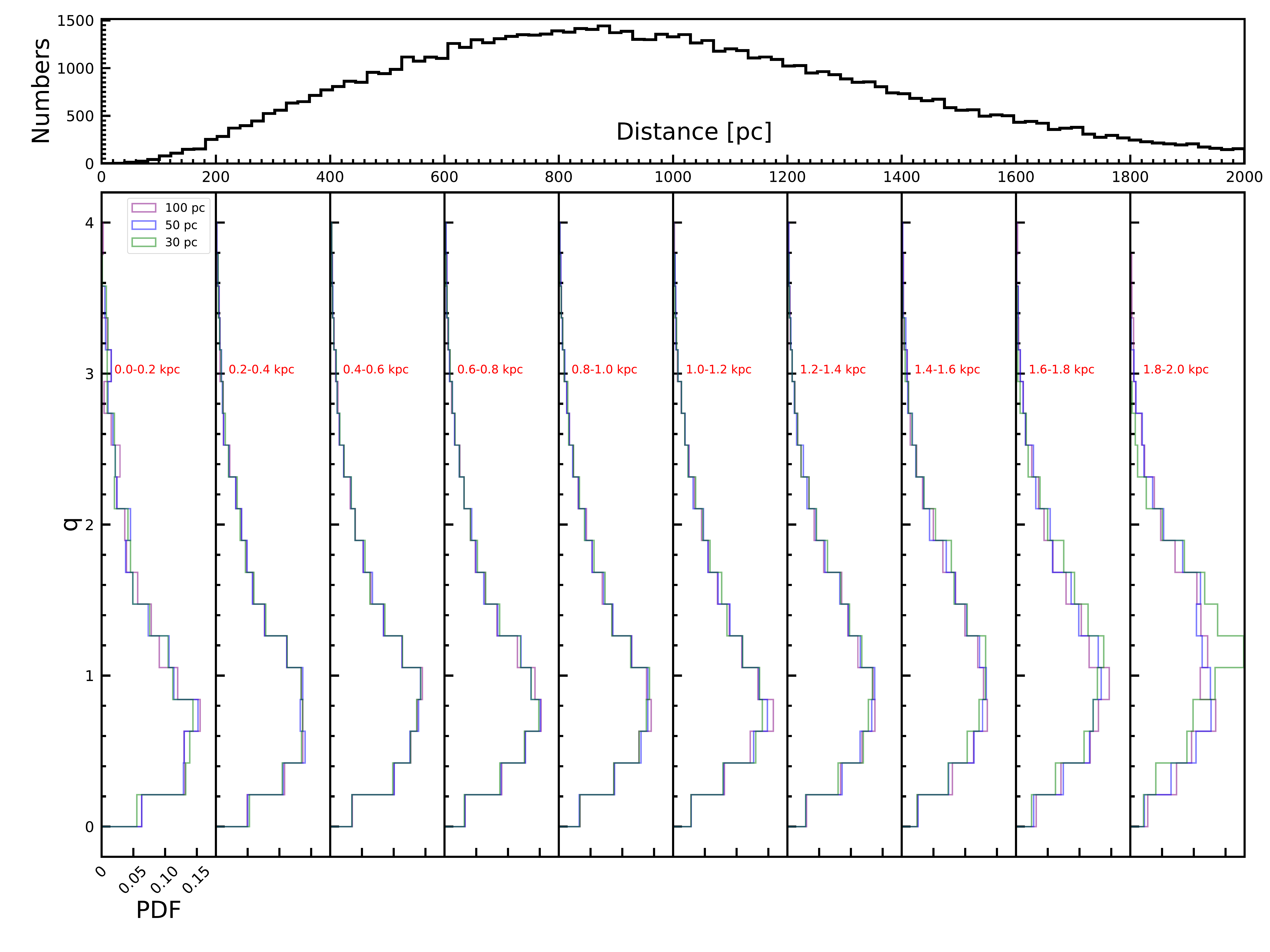}
    \caption{The distance and q-value distribution of Kepler stars. The upper figure shows the distance distribution of Kepler stars. The lower ten figures shows the q-value distribution of Kepler stars with different distances. The distance increases from left to right. Three colors (purple, blue and green) show the different ranges of stars selected to calculate the relative velocity of a given star.}
    \label{figure distance_q}
\end{figure*}

In this paper, we separate the Kepler stellar sample using the proper motions of stars rather than 3D velocities. Thus, there may be projection effects as small distances. To test the robustness of the high-, medium-, low-\textit{V} star classification correlated to different distance limits of nearby stars, we show the distance distribution and q-value distribution of Kepler stars. For a given star, we select stars within 30 pc (green), 50 pc (blue) and 100 pc (purple) of this star to calculate the relative velocity. As is shown in Figure \ref{figure distance_q}, q-value distributions of Kepler stars are nearly the same for the cases of different ranges, except for the Kepler stars with distance of 1.8--2.0 kpc. Given that the number of Kepler stars with distance of 1.8--2.0 kpc are rather small compared to the total sample, we neglect the differences due to our method of separating stars.

\section{Statistical Results}
\label{results of focc}
Many previous works have studied correlations between stellar properties and the planet occurrence rate. In this section, we focus on correlating the dependence of other stellar properties and try to obtain a more convincing correlation between the occurrence rate and the relative stellar velocities. In subsection \ref{subsub 2.2.3}, we show the correlations between stellar properties and relative velocity. In subsection \ref{sub 3.1}, we calculate the planet occurrence rate according to the methods in Appendix \ref{sub 2.3} and compare our results with previous studies to convince our validity of sample selection and calculation. Because different sized planets have different occurrence rate, we divide the whole sample into several groups in which sizes of planets are different to carefully discuss the correlations between planet occurrence rates and stellar relative velocities. In subsection \ref{sub 3.2}, we will show the occurrence rates of planets of the radius of 0.5--4 $R_{\oplus}$ around high-\textit{V} , medium-\textit{V} , and low-\textit{V} stars respectively. Because stellar properties have influence on the calculation of planet occurrence rate, in subsection \ref{sub 3.4}, we will show the results after adopting the methods of correcting the planet occurrence rate. In subsection \ref{results of multiplicity and ecc}, we will show the correlations between stellar relative velocities of planet host stars and multiplicities and average eccentricities of planetary systems.

\subsection{Correlations between stellar properties and relative velocity} \label{subsub 2.2.3}

In a total of 74002 Kepler single main-sequence stars, we find that 9754 of them are high-\textit{V} stars, 28049 of them are medium-\textit{V} stars and 36199 of them are low-\textit{V} stars. The occurrence rate of Kepler-like planets has anti-correlation with effective temperature \citep{2012ApJS..201...15H,2015ApJ...798..112M,2020AJ....159..164Y} and positive correlation with metallicity \citep{2016AJ....152..187M,2019ApJ...873....8Z}. To find a robust correlation between planet occurrence rate and stellar relative velocity, we should try our best to exclude other factors influencing the planet occurrence rate e.g. stellar effective temperature and metallicity. Therefore, we should first discuss the correlations between stellar properties and relative velocity. Here is the probability distribution function(PDF) of stellar parameters of high-\textit{V} , medium-\textit{V} , and low-\textit{V} stars. 


\begin{figure*}
    \centering
    \includegraphics[width=1\linewidth]{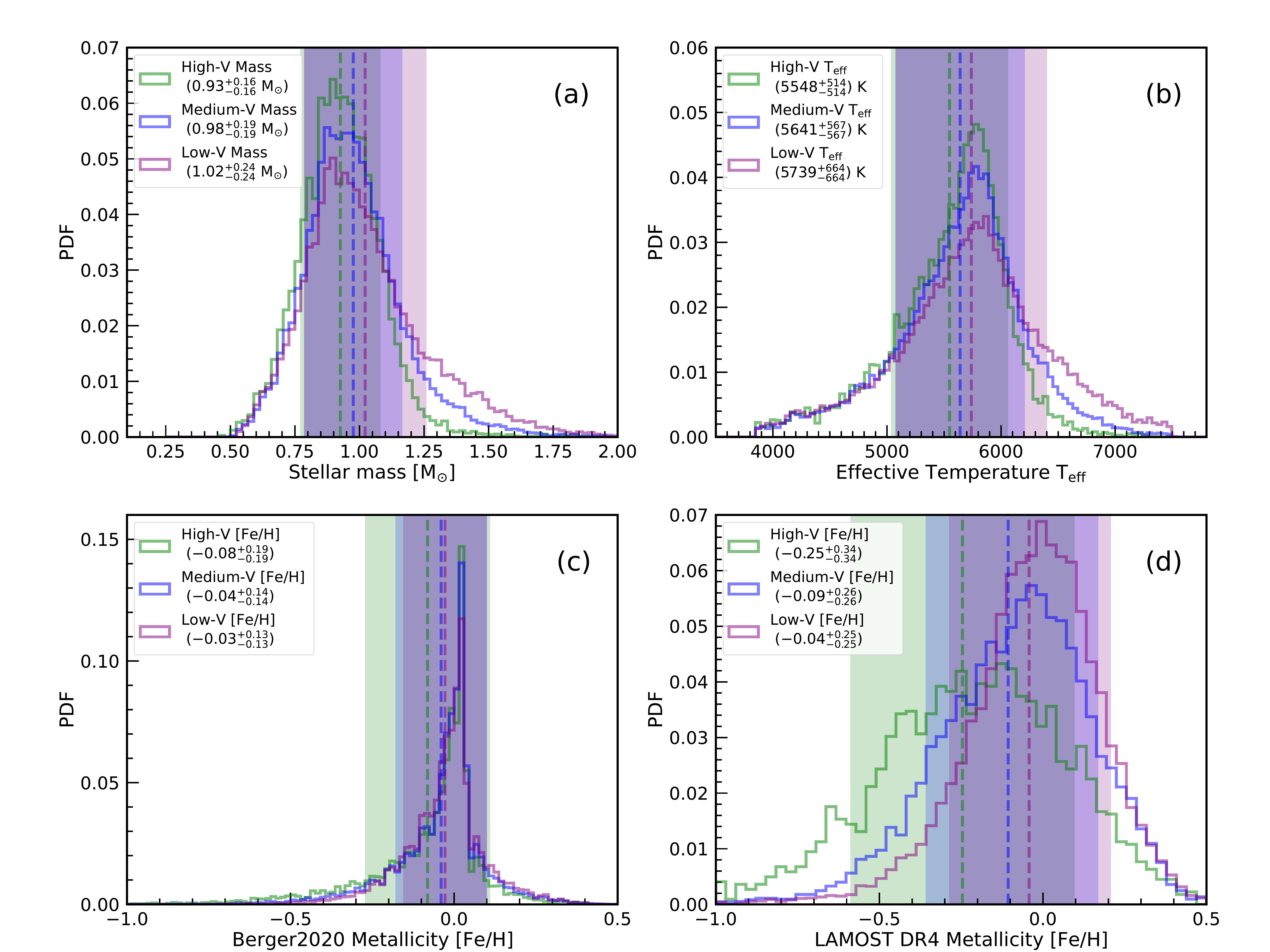}
    \caption{The distribution of stellar properties of 9754 high-\textit{V} stars, 28049 medium-\textit{V} stars, and 36199 low-\textit{V} stars. Panel (a)(b)(c)(d) show the distribution of stellar mass, effective temperature, the metallicity of B2020, and metallicity of LAMOST DR4, respectively. We use the metallicity of LAMOST DR4 from the second release of value-added catalogs of the LAMOST Spectroscopic Survey of the Galactic Anticentre (LSS-GAC DR2) \citep{2017MNRAS.467.1890X}. Green is high-\textit{V} stars; blue is medium-\textit{V} stars; purple is low-\textit{V} stars. Green, blue and purple dashed lines represent the average value of stellar properties of stars with different relative velocities. Similarly, green, blue, and purple shadow colored regions show the standard deviation of stellar properties of stars with different relative velocities.}
    \label{figure 2-4}
\end{figure*}

Fig \ref{figure 2-4} shows the distribution of stellar parameters of high-\textit{V} stars, medium-\textit{V} stars, and low-\textit{V} stars. Panel (a) shows the distribution of stellar mass. Different colors show different relative velocities. Green is high-\textit{V} stars; blue is medium-\textit{V} stars; purple is low-\textit{V} stars. Solid lines show the probability distribution function of stars. Dashed lines show the average value of stellar properties of stars with different relative velocities. Similarly, green, blue, and purple shadow colored regions show the standard deviation of stellar properties of stars with different relative velocities. The average mass of high-\textit{V} stars is 0.93$_{\rm -0.16}^{+0.16}$ M$_{\rm \odot}$ where 0.16 is the standard deviation of the stellar mass. The average mass of medium-\textit{V} stars is 0.98$_{\rm -0.19}^{+0.19}$ M$_{\rm \odot}$. The average mass of low-\textit{V} stars is 1.02$_{\rm -0.24}^{+0.24}$ M$_{\rm \odot}$. Considering the average uncertainty in stellar mass is relatively small, about 7\%, we can roughly conclude that the average mass of stars tends to decrease with increasing relative velocity. Or we can say that stars with lower mass prefer to move with higher relative velocity. Similarly, in panel (b)---the distribution of effective temperature, we can also conclude that stars with higher relative velocity prefer to have a lower effective temperature. More specifically, with the consideration of 3\% $T_{\rm eff}$ errors or $\sim 112$ K, the average effective temperature of high-\textit{V} stars is higher than that of low-\textit{V} stars with $\sim$ 1.5 $\sigma$. This is easy to understand, because of the strong positive correlation between stellar mass and effective temperature coming from the well-known empirical mass-luminosity relation. Because some previous study argued the planet occurrence rate increases with decreasing stellar mass or stellar effective temperature \citep{2012ApJS..201...15H,2015ApJ...798..112M,2020AJ....159..164Y}. Therefore, we should discuss the potential influence of stellar mass or effective temperature in the following calculation of the planet occurrence rate.

Panel (c) shows the distribution of stellar metallicity calculated in B2020. The average metallicity of high-\textit{V} stars is -0.08$_{\rm -0.18}^{+0.18}$, that of medium-\textit{V} stars is 0.04$_{\rm -0.14}^{+0.14}$ and that of low-\textit{V} stars is 0.03$_{\rm -0.13}^{+0.13}$. Although the average metallicity decreases slightly with increasing stellar relative velocity, considering the relatively large standard error of the metallicity, stars with different relative velocities have similar stellar metallicity 
based on data from B2020 (with differences of 0.4 $\sigma$). 

Since \cite{2014ApJ...789L...3D} have pointed out that the metallicity from Kepler star \citep{2014ApJS..211....2H} has significant systematic biases. To check whether the metallicity calculated in \cite{2020AJ....159..280B} have systematic biases, we choose the data of LAMOST as a comparison with caution. Since LAMOST is initially designed for the stellar spectral survey in the Milky Way, it's ideal for measuring the metallicity of stars in the Kepler region, although LAMOST does not cover the entire sample of Kepler stars. After cross-matching LAMOST DR4 and our selected stellar samples, we find $\sim$ 22000 stars (28\%) having metallicity of LAMOST DR4. There are about 2219 high-\textit{V} stars, 7294 medium-\textit{V} stars, and 11377 low-\textit{V} stars among these stars. If stars in the cross-matched catalog have significantly different fractions of stars with different relative velocities from our selected stellar samples, it suggests that high-\textit{V} , medium-\textit{V} , and low-\textit{V} stars are not homogeneously distributed in the Kepler region. Therefore, before using the Kepler LAMOST cross-matched data, we should first check the fraction of stars with different relative velocities both in Kepler LAMOST cross-matched catalog and our selected stellar catalog. We find that they have similar fractions in different catalogs i.e. high-\textit{V} stars: 13\% (B2020) vs 11\% (LAMOST DR4),  medium-\textit{V} stars: 38\% (B2020) vs 35\% (LAMOST DR4), and low-\textit{V} stars: 49\% (B2020) vs 54\% (LAMOST DR4). Therefore, we can roughly conclude that stars with different relative velocities are homogeneously distributed.  We can use metallicity obtained in LAMOST DR4 to represent the whole main-sequence stars in the Kepler region.  

Panel (d) shows the distribution of stellar metallicity obtained from LAMOST DR4. The average metallicity of high-\textit{V} stars is -0.25$_{\rm -0.34}^{+0.34}$, that of medium-\textit{V} stars is -0.09$_{\rm -0.26}^{+0.26}$ and that of low-\textit{V} stars is 0.04$_{\rm -0.25}^{+0.25}$. The standard error of metallicity in LAMOST DR4 is about 0.1 dex. Similar to the analysis above, we can easily get the result that high-\textit{V} stars prefer lower metallicity than low-\textit{V} stars 
with $\sim$ 1.7 $\sigma$ differences.

The correlation between stellar metallicity and relative velocity shows significantly different trends in panel (c) and (d)(more details in Appendix \ref{app A}). The difference may be attributed to those stars without  constraints of spectroscopic metallicity. In the following subsections, we will both discuss the influence of metallicity from different tables(metallicity of B2020 and metallicity of LAMOST DR4) on correlations between planet occurrence rate and stellar relative velocity.

To sum up, we find some correlations between stellar properties and stellar relative velocity, i.e. high-\textit{V} stars with higher relative velocity, have smaller average stellar mass, lower average effective temperature, and lower average stellar metallicity(LAMOST DR4).

\subsection{Comparison with previous results}
\label{sub 3.1}
\begin{figure*}
    \centering
    \includegraphics[width=0.95\linewidth]{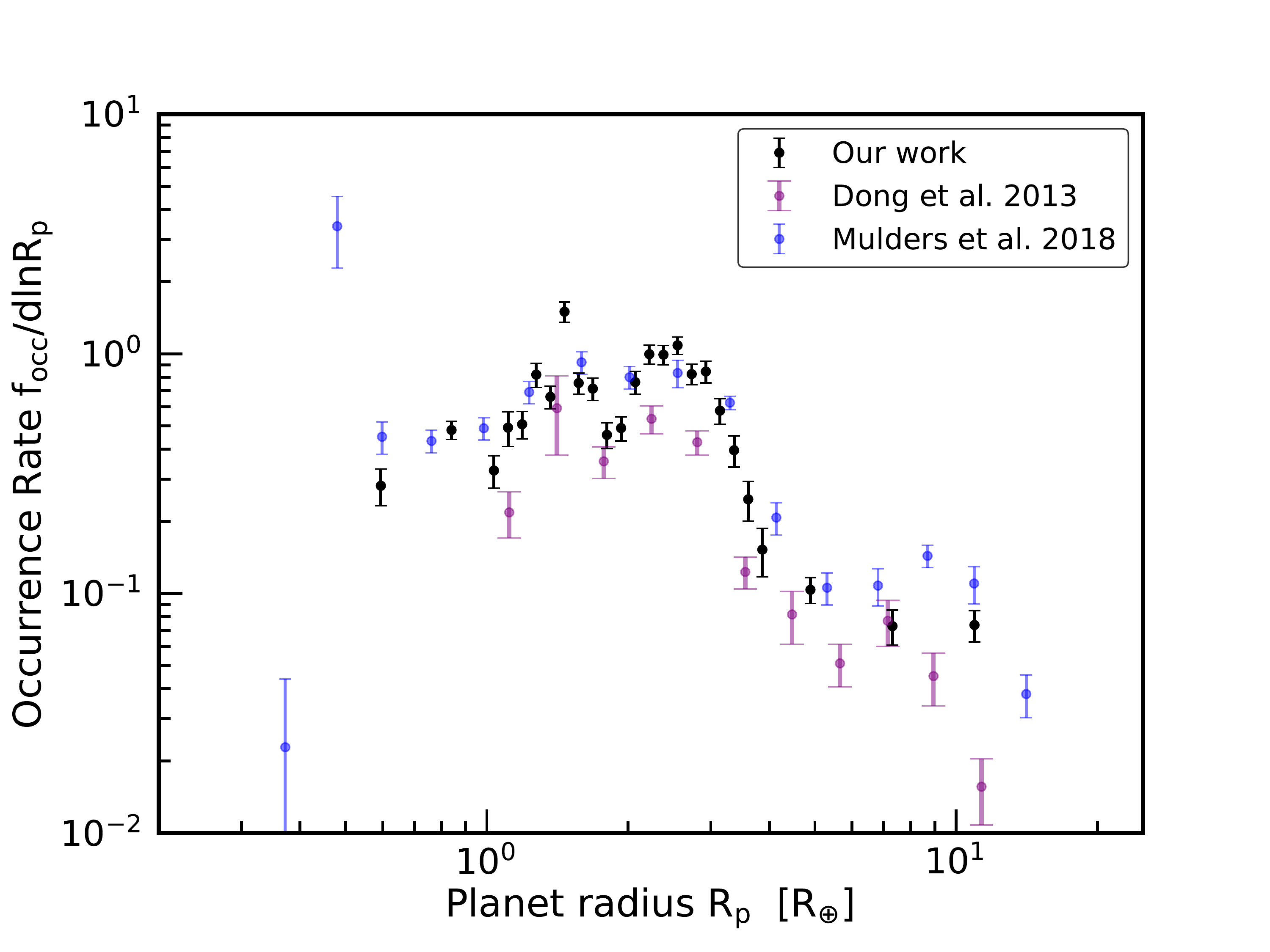}
    \caption{Relation between normalized planet occurrence rate(f$_{\rm  occ}$/dlnR$_{\rm p}$) and planet radius(R$_{\rm p}$). The black bars are show the occurrence rates estimated using the method in appendix C. The green, blue, and purple bars shows the results calculated in \cite{2013ApJ...778...53D} and \cite{2018AJ....156...24M}.}
    \label{figure 3-1}
\end{figure*}

Previous work shows the planet occurrence rate both correlated with planet radius and orbital periods. In this section, we choose the planet with periods less than 400 days to estimate the occurrence rate. Furthermore, to minimize the influence of the size of bins, we calculate the normalized planet occurrence rate (i.e. f$_{\rm occ}$/dlnR$_{\rm p}$, the number of planets per star normalized in different bin size in lnR$_{\rm p}$ space).

Fig \ref{figure 3-1} shows the relation between normalized planet occurrence rate and planet radius(R$_{\rm p}$). Our planet occurrence rates are nearly the same as those in \cite{2018AJ....156...24M}, but slightly higher than the planet occurrence rates calculated by \cite{2013ApJ...778...53D}. Our sample includes 74002 stars and 1910 planets, while the sample used by \cite{2013ApJ...778...53D} includes $\sim$ 120000 stars and 2300 planet candidates. Thus, our sample will have a slightly higher of average planet numbers per star i.e. (1910/74002)/(2300/120000) $\sim$ 1.35.  

We find the bi-modal structure at a range of 1--4 $R_{\rm \oplus}$ in the distribution planet of planet radius and planet occurrence rate which is highly consistent with that well-known 1.8 $R_{\oplus}$ planet radius gap found by \cite{2017AJ....154..109F}. After the second peak around $\sim$ 2.5--3 $R_{\oplus}$, normalized planet occurrence rate declines rapidly with the increasing planet radius. This rapid declination of planet occurrence rate may associate with planet desert due to the radiation of host stars. Additionally, there is an ambiguous plateau in the range of $\sim$ 4--10 $R_{\oplus}$. Although there are only several points, both our work(black dots), \cite{2013ApJ...778...53D}(purple dots), and \cite{2018AJ....156...24M}(blue dots) show a similar relation. The plateau indicates the different characteristics of giant planets and Kepler-like planets including super-Earth and sub-Neptune. Kepler has not detected many planet candidates in some specific ranges of orbital periods, especially for cold gas giants. The occurrence rates of these planets with longer periods are underestimated significantly. Gaia spacecraft may detect more cold Jupiters with long periods through the astrometric method. More cold Jupiters can be used to check whether there is a plateau for gas giants in the distribution plane of planet occurrence rate and planet radius, and improve our knowledge of the formation and evolution of cold Jupiters.

\subsection{occurrence rates of planets with 0.5-4 $R_{\oplus}$} \label{sub 3.2}



\begin{figure*}     \centering
    \includegraphics[width=1\linewidth]{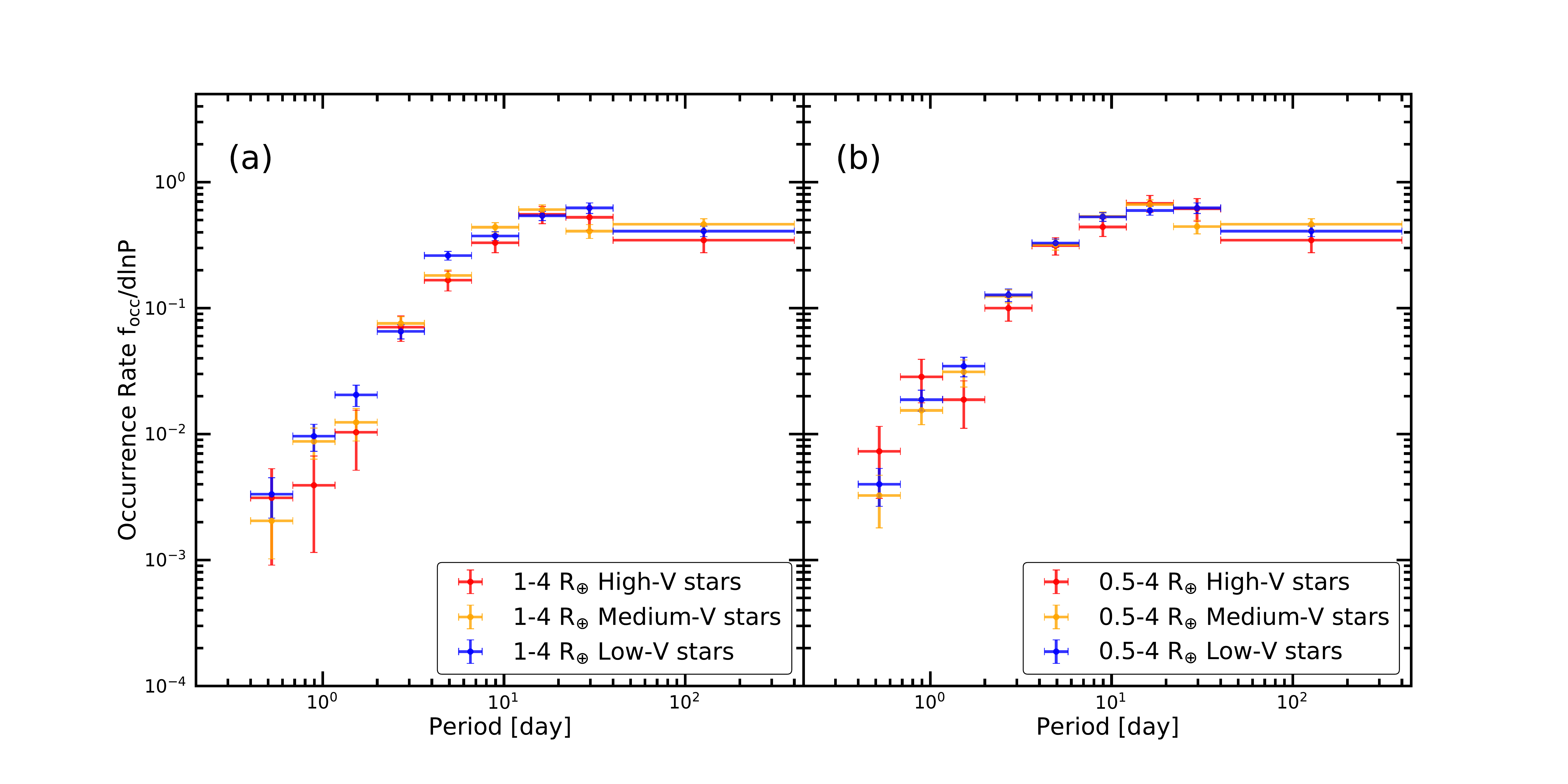}
    \caption{The relation between normalized planet occurrence rate($f_{\rm occ}$/lnP) and stellar relative velocity in the distribution of planet orbital period. The red, yellow, and blue symbols show the planets around stars with high, median, and normal relative velocity respectively. Panel (a) shows relation for planets with radius of 1-4 $R_{\oplus}$; Panel (b) shows relation for planets with radius of 0.5-4 $R_{\oplus}$.}    \label{figure 3-2}
\end{figure*}

To investigate the correlation between occurrence rate and planet orbital period, we choose the planets with a radius of 1--4 $R_{\oplus}$ and 0.5--4 $R_{\oplus}$. We calculate the normalized planet occurrence rate in a different range of orbital periods(f$_{\rm  occ}$/dlnP). In Fig \ref{figure 3-2}, we show the relation between planet occurrence rates and stellar relative velocity. Panel (a) shows the results for planets with radius of 1--4 $R_{\oplus}$ and Panel (b) shows the results for planets with radius of 0.5--4 $R_{\oplus}$. In both panels, no matter the planet hosts are  high-\textit{V} , medium-\textit{V} , or low-\textit{V} stars, the planet occurrence rate increases with increasing orbital periods within the 10 days ($\sim$ 0.1 au for solar-mass star). When the orbital periods 10 $\textless$ P $\textless$ 400 days, the planet occurrence rate keeps constant as a plateau. 

The broken law of occurrence rate is consistent with \cite{2012ApJS..201...15H,2013ApJ...778...53D,2015ApJ...798..112M}. Several mechanisms may interpret the break around 10 days. The break-in planet occurrence rate at $\sim$ 10 days can be attributed to the truncation of protoplanetary disks by their host star magnetospheres at co-rotation radius \cite{2015ApJ...798..112M,2017ApJ...842...40L}. Because under the assumption of in-situ formation scenario, lower solid material left in the protoplanetary disks within 10 days consequently result in the lower planet occurrence rate. Additionally, for solar-mass stars, orbital period P $\sim$ 10 days i.e. semi-major axis a $\sim$ 0.1 au, near the location of sublimation front of silicate \cite{2019A&A...630A.147F}. Both in-situ formation and migration scenario can explain this break at $\sim$ 10 days, i.e. for inward-drifting pebbles to accumulate and form planets at the pressure maximum a short distance outside the dust sublimation front \cite{2019A&A...630A.147F} or for inward migration of multiple planets where the first planet is trapped at the inner disk edge and halts the migration of other planets (\cite{2014A&A...569A..56C}). 

Besides the broken law, we find interesting correlations between planet occurrence rates and stellar relative velocity. In panel (a), planets with the radius of 1--4 $R_{\oplus}$ around high-\textit{V} stars have a lower occurrence rate than that around low-\textit{V} stars with 0.85 $\sigma$ differences on average. In panel (b), planets within 1 days with the radius of 0.5--4 $R_{\oplus}$ around high-\textit{V} stars have a slightly higher occurrence rate compared with planets around low-\textit{V} stars. While for planets outside 1 days, planets around high-\textit{V} stars still have a lower occurrence rate on average, i.e. with 0.59 $\sigma$ difference compared with planets around low-\textit{V} stars. The difference of occurrence rate planets within 1 days in panel (a) and (b) indicates that short-period sub-Earth-sized planets (0.5--1 $R_{\oplus}$) prefer to orbit around high-\textit{V} stars. 

\begin{figure*} 
    \centering
    \includegraphics[width=1\linewidth]{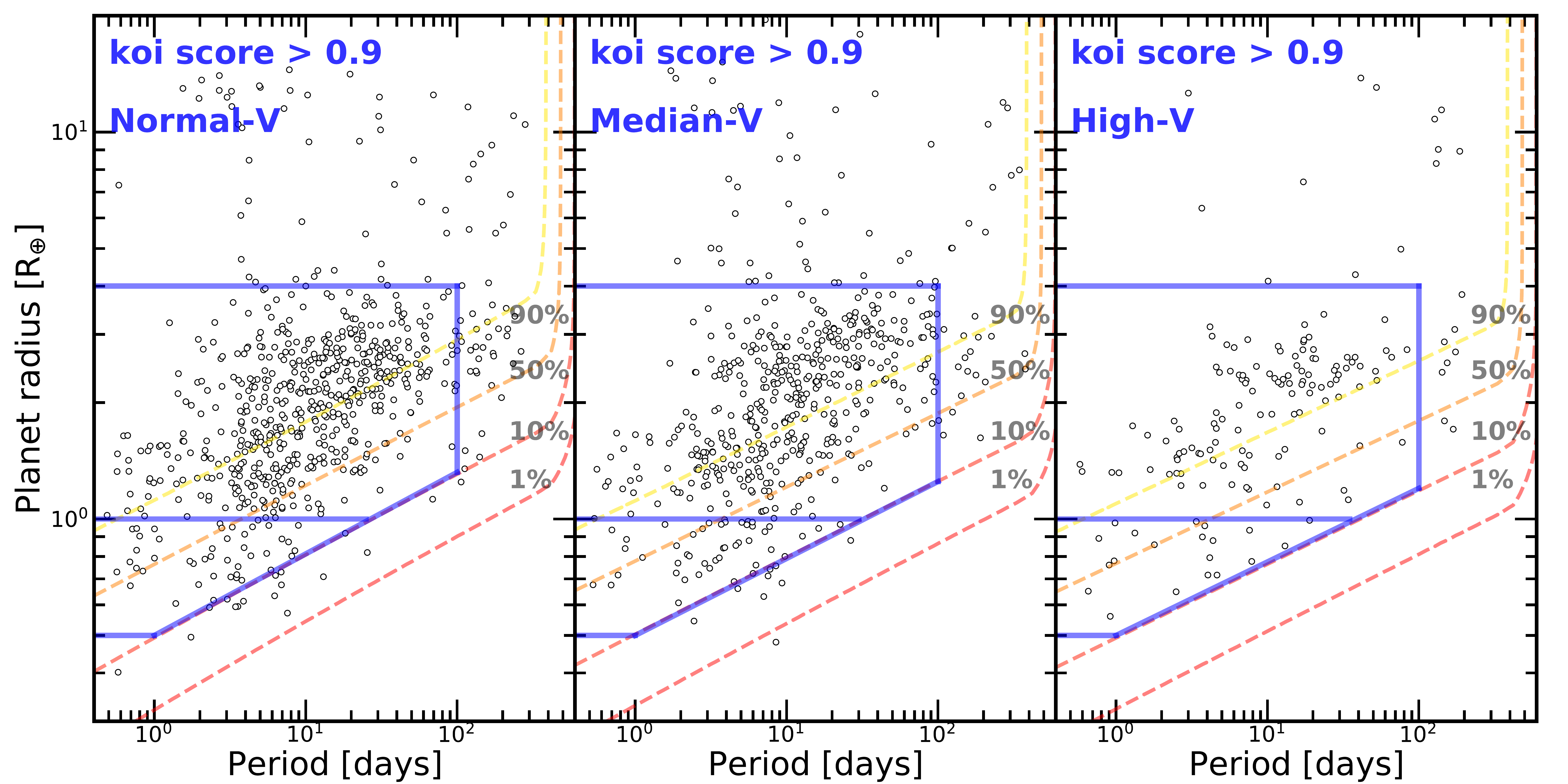}
    \caption{The orbital period and planet radius distribution. Here we select the planets with disposition score $\textgreater$ 0.9. The upper green PDF shows the distribution of the orbital period, while the right red PDF shows the distribution of the planet radius. The contour lines(yellow to red) in the mid panel, represents the detection completeness $\eta$ (for a planet with given planet radius and orbital periods, the completeness means the the number of Kepler main-sequence single stars that can detect the planet over the number of Kepler main-sequence single stars in different V bins) of 90\%, 50\%, 10\%, 1\%. You can also find the definition of completeness in Appendix B}
    \label{figure 3-3}
\end{figure*}
\begin{figure*} 
    \centering
    \includegraphics[width=1\linewidth]{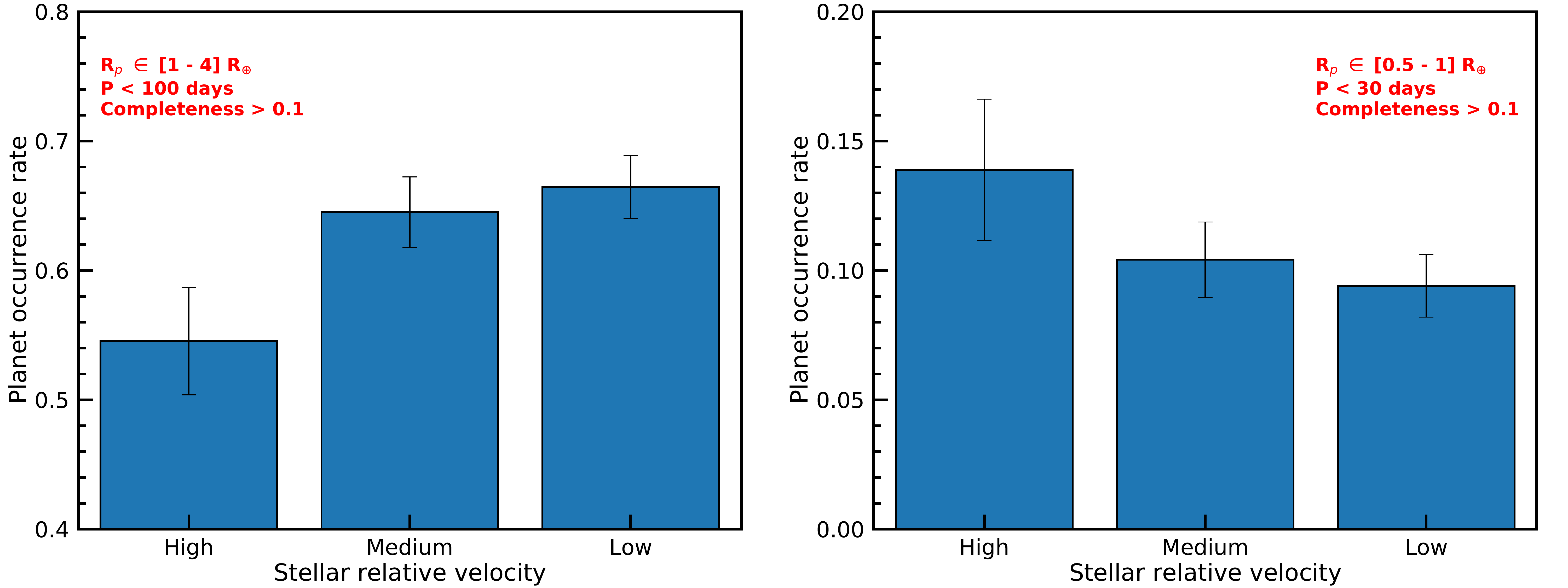}
    \caption{Occurrence rate of planets around high-\textit{V} , medium-\textit{V} and low-\textit{V} stars. The left panel shows the result of super-earth- and sub-neptune- sized planets(1--4 $R_{\oplus}$, P$\textless$100 days). The right panel shows the result of sub-earth-sized planets(0.5--1 $R_{\oplus}$, P$\textless$ 30 days)}
    \label{figure 3-4}
\end{figure*}

Here we define detection completeness $\eta$ of a given planet as the fraction of stars that the planet can be detected around (more details in Appendix B). In our calculation, for planets with low detection completeness, small $\eta$ will result in large uncertainties of planet occurrence rate with the assumption of Poisson distribution. To minimize the influence of planets with low detection completeness, we set another parameter cutting criterion, i.e. selecting planets with completeness $\eta$ $>$ 0.1. Fig \ref{figure 3-3} shows the detection completeness of highly reliable planet candidates whose $Robovetter$ disposition scores are larger than 0.9 in the distribution plane of planet radius and orbital period for stars with different relative velocities. For planets with radius of 1--4 $R_{\oplus}$, we select planets with period P $\textless$ 100 days. While for planets with radius of 0.5--1 $R_{\oplus}$, we both select planets within 30 days. Fig \ref{figure 3-4} shows the occurrence rate of the planet around high-\textit{V} , medium-\textit{V} , and low-\textit{V} stars with cut of detection completeness, i.e. orbital periods cut. For super-Earth with the radius of 1--4 $R_{\oplus}$ around high-\textit{V} stars, their occurrence rate is significantly lower than the occurrence rate of planets around low-\textit{V} stars, i.e. with 2.5 $\sigma$ differences. For sub-Earth with the radius of 0.5--1 $R_{\oplus}$ around high-\textit{V} stars, the occurrence rate is slightly higher (1.5 $\sigma$) than the occurrence rate of planets around low-\textit{V} stars. These results are consistent with that in Fig \ref{figure 3-2}. However, we don't exclude the influences of stellar properties such as stellar mass and metallicity on the planet occurrence rate i.e. planet occurrence rate has anti-correlation with effective temperature \citep{2012ApJS..201...15H,2015ApJ...798..112M,2020AJ....159..164Y} and positive correlation with metallicity (\cite{2016AJ....152..187M,2019ApJ...873....8Z}. In Appendix \ref{sub 3.3}, we introduce two methods to minimize the influence of stellar properties, i.e. prior correction and posterior corrections.

\subsection{Planet occurrence rate after corrections} \label{sub 3.4}

\begin{figure}
    \centering
    \includegraphics[width=1\linewidth]{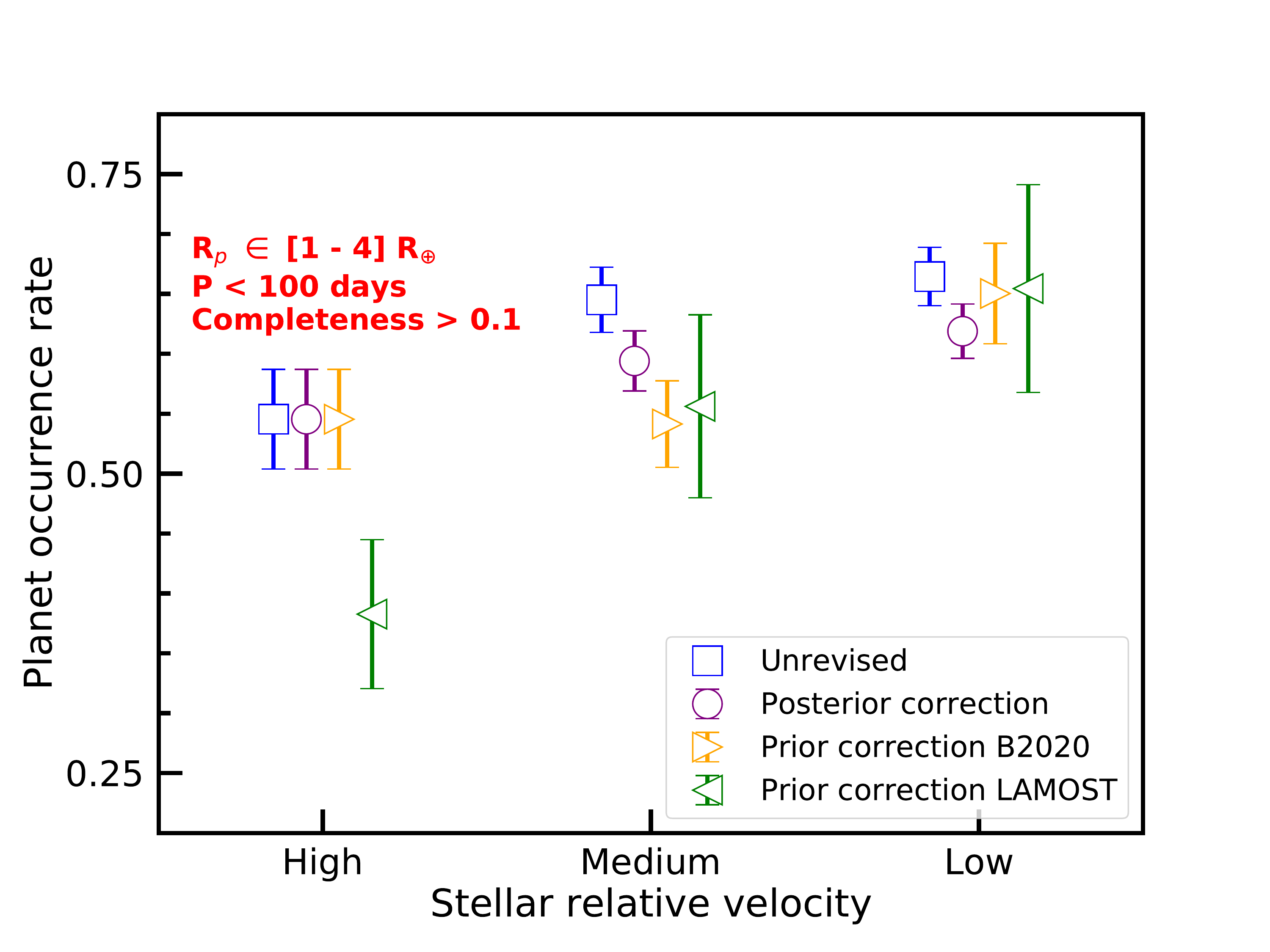}
    \caption{The correlation between the planet occurrence rate and stellar relative velocity. We shows the results of planets with radius $R_{\rm p}$ $\in$ 1--4 $R_{\oplus}$ and orbital period P $\textless$ 100 days. Blue squares show the unrevised planet occurrence rate. Purple circles show the planet occurrence rate after posterior correction. Orange triangles show the planet occurrence rate after prior correction utilizing metallicity in B2020. Green triangles show the planet occurrence rate after prior correction LAMOST DR4.}
    \label{figure 3-5}
\end{figure}

\begin{deluxetable*}{lccc}
\tablecaption{Values in Fig. \ref{figure 3-5}\label{table 2}}
\tablewidth{0pt}
\tablehead{
\colhead{Case} &\colhead{high-\textit{V} } stars &\colhead{medium-\textit{V} } stars &\colhead{low-\textit{V} } stars
} 
\startdata
Unrevised, $f_{\rm occ}$ &0.545 &0.645 &0.665 \\
Unrevised, $\sigma_{\rm f_{\rm occ}}$ &0.042 &0.027 &0.024 \\
Unrevised, differences &  & 2.47 $\sigma$ &  \\
\hline
Posterior correction, $f_{\rm occ}$ &0.545 &0.594 &0.619 \\
Posterior correction, $\sigma_{\rm f_{\rm occ}}$ &0.042 &0.025 &0.023 \\
Posterior correction, differences & & 1.55 $\sigma$ & \\
\hline
Prior correction B2020, $f_{\rm occ}$  &0.545 &0.541 &0.650 \\
Prior correction B2020, $\sigma_{\rm f_{\rm occ}}$&0.042 &0.036 &0.042 \\
Prior correction B2020, differences & &1.78 $\sigma$& \\
\hline
Prior correction LAMOST, $f_{\rm occ}$ &0.383 &0.556 &0.654 \\
Prior correction LAMOST, $\sigma_{\rm f_{\rm occ}}$&0.062 &0.076 &0.087 \\
Prior correction LAMOST, differences & &2.55 $\sigma$& \\
\hline
Prior correction LAMOST + Posterior correction, $f_{\rm occ}$ &0.383 &0.512 &0.610 \\
Prior correction LAMOST + Posterior correction, $\sigma_{\rm f_{\rm occ}}$&0.062 &0.070 &0.081 \\
Prior correction LAMOST + differences, confidence level & &2.23 $\sigma$& \\
\enddata
\tablecomments{we list the values of planet occurrence rate($f_{\rm occ}$) and its errors($\sigma_{\rm f_{\rm occ}}$). The confidence level is defined as the difference between the occurrence rate of planets around high-\textit{V} and low-\textit{V} stars. Here we not only list the values of cases named "unrevised", "Posterior correction", "Prior correction B2020" and "Prior correction LAMOST", but also add the case named "Prior correction LAMOST + Posterior correction". As mentioned in subsection \ref{sub 3.4}, for the case of "Prior correction LAMOST", because we are unable to select the nearest samples passing KS two-sample test i.e. P $\textgreater$ 0.05, we combine the Prior correction and Posterior correction to get a convincing value.}
\end{deluxetable*}

We both use posterior correction and prior correction to exclude the influence of stellar properties on the planet occurrence rate and confirm whether stellar relative velocity is another factor correlated with the planet occurrence rate. Because of the difference between metallicity in B2020 and LAMOST DR4, we use both of the metallicities in B2020 and LAMOST DR4 to do the prior correction respectively, i.e. Prior correction B2020 and Prior correction LAMOST. Because the empirical correlations between planet occurrence rate and stellar properties are different for planets with different sizes. We will show the results after the correction of the occurrence rate of planets with different radius, i.e. correction of the occurrence rate of planets with the radius of 1--4 $R_{\oplus}$, 0.5--1 $R_{\oplus}$ and 4--20 $R_{\oplus}$.

\subsubsection{Correction of occurrence rate of planets with radius of 1--4 $R_{\oplus}$} \label{cor 1-4}

\begin{figure*}
    \centering
    \includegraphics[width=0.95\linewidth]{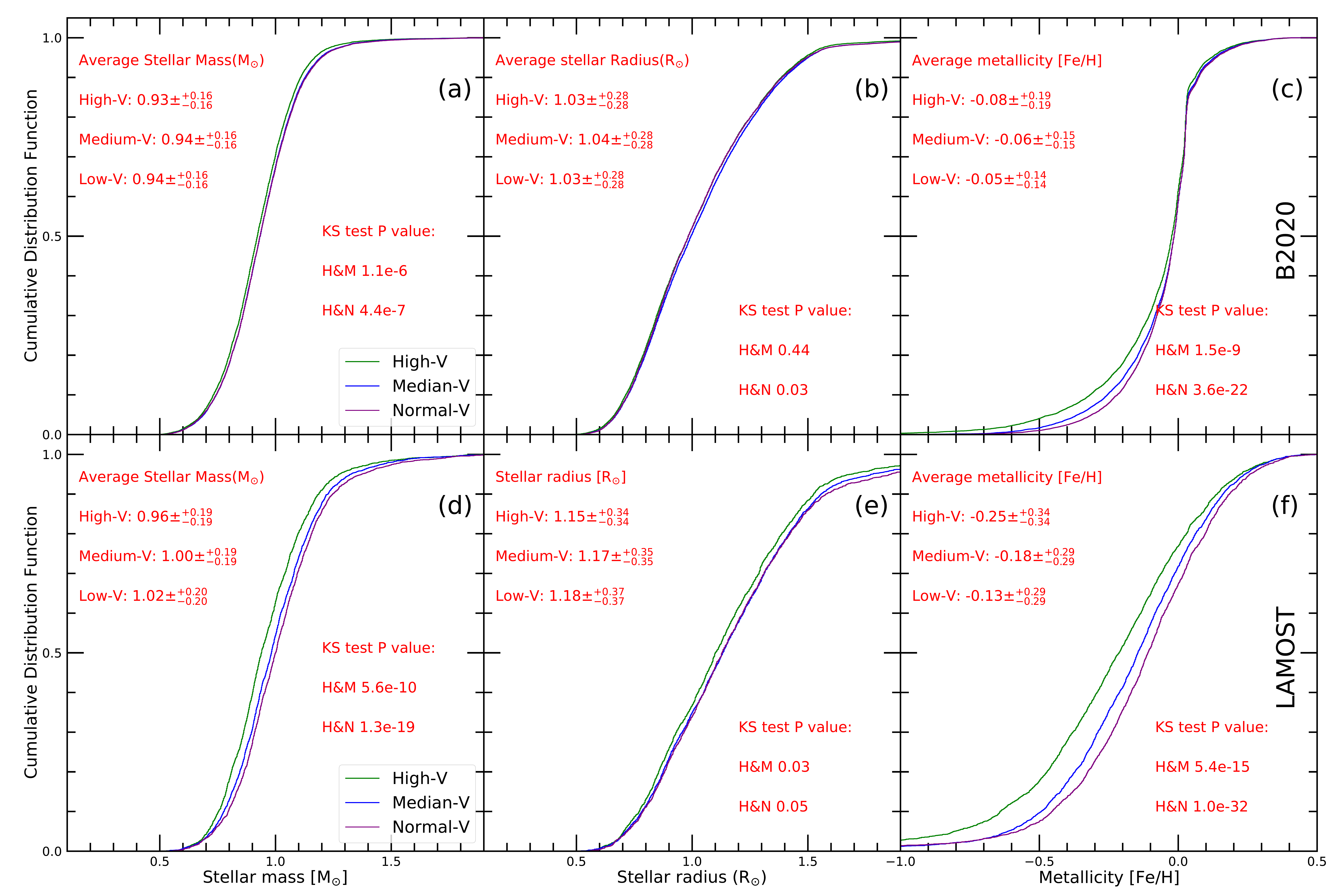}
    \caption{The cumulative distribution function(CDF) of stellar mass(panel (a)), radius(panel (b)) and metallicity of B2020(panel (c)), for high-\textit{V} , medium-\textit{V} and low-\textit{V} stars respectively. We select medium-\textit{V} and low-\textit{V} stars with the nearest stellar properties of every given high-\textit{V} star. Green, blue and purple shows high-\textit{V} stars and those selected medium-\textit{V} and low-\textit{V} stars respectively. In each panel, we also list the average values of stellar properties and p values of the KS two-sample test. H\&M means KS two-sample test between high-\textit{V} and those selected medium-\textit{V} stars. H\&N means KS two-sample test between high-\textit{V} and those selected low-\textit{V} stars. Panels (d)(e)(f) are similar to Panels (a)(b)(c). The difference lies in the metallicity, i.e. panels (d)(e)(f) use stars with metallicity of LAMOST DR4, while panels (a)(b)(c) use stars with metallicity of B2020.}
    \label{figure B-2}
\end{figure*}
In Figure \ref{figure 3-5}, we show the result for planets with a radius of 1–4 $R_{\oplus}$ after correction. Different colors show the results of different cases. Blue squares show the unrevised planet occurrence rate. Purple circles show the planet occurrence rate after posterior correction. Orange and green triangles show the planet occurrence rate after prior correction utilizing the metallicity of B2020 and prior correction utilizing the metallicity of LAMOST DR4, respectively. All these results show an anticorrelation between stellar relative velocity and occurrence rate of planets with a radius of 1–4 $R_{\oplus}$.

In Posterior correction, we use empirical $f_{\rm occ}-T_{\rm eff}$ and $f_{\rm occ}-[Fe/H]$ correlations to correct the influence of $T_{\rm eff}$ and $[Fe/H]$. Here stellar effective temperatures are taken from B2020 and metallicities are taken from LAMOST DR4. medium-\textit{V} and low-\textit{V} stars have originally higher average $T_{\rm eff}$ and lower average metallicity than high-\textit{V} stars. In order to correct the planet occurrence rate due to different distributions of stellar properties, we use efficiencies calculated with Equation (\ref{equation 3-5}) and (\ref{equation 3-6}), i.e. $C_{\rm medium \Rightarrow high}$ = 0.92 and $C_{\rm low \Rightarrow high}$ = 0.93. Before the posterior correction, the difference in the occurrence rate of the planet around high-\textit{V} and low-\textit{V} stars is about 2.47 $\sigma$. After correction, the difference is about 1.55 $\sigma$, which indicates that stellar relative velocity is likely another factor influencing the planet occurrence rate. In the metallicity correction, we use the data from figure 3 in \cite{2019ApJ...873....8Z}. Because the data of figure 3 in \cite{2019ApJ...873....8Z} includes large planets, the empirical Equation (\ref{equation 3-4}) overestimates the positive correlations between planet occurrence rate and metallicity. If using the data excluding large planets, the $f_{\rm occ}-[Fe/H]$ correlation is even weaker. Consequently, the difference in occurrence rates of planets around high-\textit{V} and low-\textit{V} stars after posterior correction will be a little bit larger than 1.55 $\sigma$.

In the Prior correction B2020, we use the NearestNeighbors function in scikit-learn \cite{JMLR:v12:pedregosa11a} to select 12856 medium-\textit{V} and 12908 low-\textit{V} stars whose stellar properties are the nearest to that of 9754 high-\textit{V} stars. We also select 225 planets with radius of 1--4 $R_{\oplus}$ around medium-\textit{V} stars, 241 planets with radius of 1--4 $R_{\oplus}$ around low-\textit{V} stars, and 38 planets with radius of 1--4 $R_{\oplus}$ around high-\textit{V} stars correspondingly. In Fig \ref{figure B-2}, the panel (a)(b)(c) show the cumulative distribution function(CDF) of stellar mass, stellar radius and metallicity of B2020 respectively. Although K-S two sample tests shows low P value(i.e. $P_{\rm KS} \leqslant 0.05$), the D value is small(i.e. the largest difference between the two cumulative distribution function is lower than 5\%). And the little difference of average stellar mass, stellar radius and metallicity (lower than 5\%) indicate that their distributions are statistically similar(i.e. even if add another Posterior correction, the results will be nearly the same). After Prior correction B2020, the difference of occurrence rates of planets around high-\textit{V} and low-\textit{V} stars is about 1.78 $\sigma$. 

In the Prior correction LAMOST, we only select 2835 medium-\textit{V} and 2621 low-\textit{V} stars whose stellar properties are the nearest to that of 2219 high-\textit{V} stars. We also select 53 planets with radius of 1--4 $R_{\oplus}$ around medium-\textit{V} stars, 57 planets with radius of 1--4 $R_{\oplus}$ around low-\textit{V} stars, and 38 planets with radius of 1--4 $R_{\oplus}$ around high-\textit{V} stars correspondingly. Although due to limited numbers of stars and planets, the occurrence rates of planets after Prior correction LAMOST have relatively large uncertainty, the difference in occurrence rates of planets around high-\textit{V} and low-\textit{V} stars is significant(2.55 $\sigma$). We use Nearest-Neighbors to select the stars with similar metallicity, however, these three stellar groups show significant difference in the distribution of metallicity, although is smaller than that in Fig \ref{figure 2-4}(panel (d)(e)(f) in Fig \ref{figure B-2}). If we add another Posterior correction, the difference declines to 2.23 $\sigma$. You can find values in table \ref{table 2}.  

Both of these occurrence rates of planets ($R_{\rm p}$ $\in$ 1--4 $R_{\oplus}$, P $\textless$ 100 days) with and without correction show the anti-correlation between planet occurrence rate and stellar relative velocity, which indicates that stellar relative velocity is likely another factor influencing planet occurrence rate. For posterior correction and prior correction B2020, the difference of occurrence rates of planets around high-\textit{V} and low-\textit{V} stars is 1.78 $\sigma$, while for prior correction LAMOST, the difference is even more significant(2.55 $\sigma$). Different ways of correction show results with a similar trend. Therefore, the anti-correlation between planet occurrence rates and stellar relative velocities is relatively robust.


\subsubsection{Correction of occurrence rate of planets with radius of 0.5--1 $R_{\oplus}$} \label{cor 0.5-1}
\begin{figure}
    \centering
    \includegraphics[width=1\linewidth]{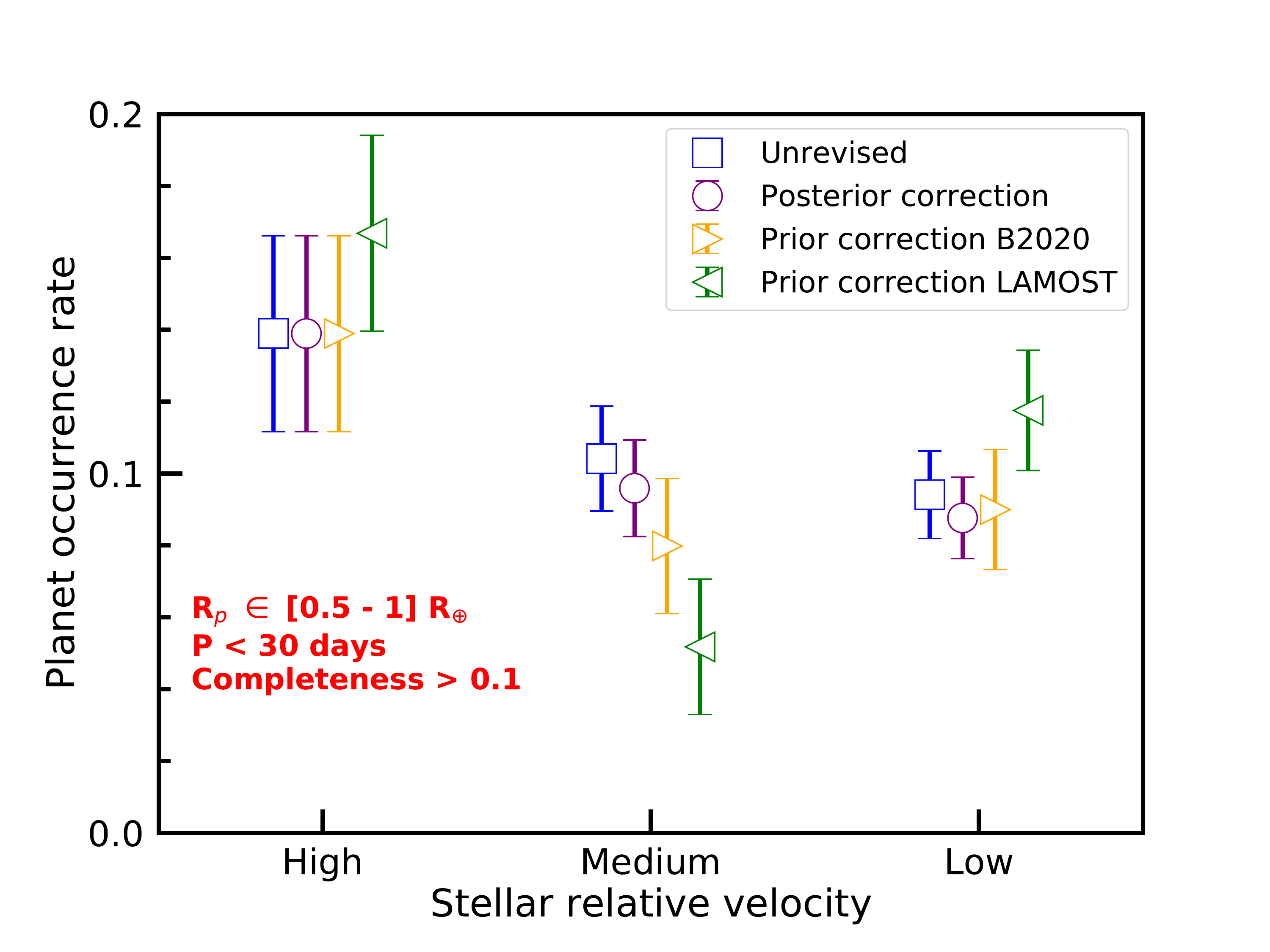}
    \caption{The correlation between the planet occurrence rate and stellar relative velocity. We shows the results of planets with radius $R_{\rm p}$ $\in$ 0.5--1 $R_{\oplus}$ and orbital period P $\textless$ 30 days. Blue squares show the unrevised planet occurrence rate. Purple circles show the planet occurrence rate after posterior correction. Orange triangles show the planet occurrence rate after prior correction utilizing metallicity in B2020.}
    \label{figure 3-6}
\end{figure}

In addition to the correction of the occurrence rate of planets with a radius of 1--4 $R_{\oplus}$, we also correct the occurrence rate of planets with a radius of 0.5--1 $R_{\oplus}$. In Fig \ref{figure 3-6}, blue squares show the unrevised planet occurrence rate. Purple circles show the planet occurrence rate after posterior correction. Orange and green triangles show the planet occurrence rate after prior correction utilizing the metallicity of B2020 and prior correction utilizing the metallicity of LAMOST DR4 respectively. 

In the posterior correction(purple circles in Fig \ref{figure 3-6}), we use the same methods and same selection samples in Fig \ref{figure 3-5}. After correction, the difference of occurrence rates of planets around high-\textit{V} and low-\textit{V} stars is 1.73 $\sigma$, which is slightly higher than the differences in the case with unrevised data(1.5 $\sigma$).

In the prior correction B2020, we also select 12908 medium-\textit{V} and 12865 low-\textit{V} stars(planets with the same ranges of parameters as that described in figure \ref{figure 3-6}) whose stellar properties are the nearest to that of 9765 high-\textit{V} stars. After correction, the difference in occurrence rates of planets around high-\textit{V} and low-\textit{V} stars is 1.53 $\sigma$ which is similar to the case of unrevised data.

Similarly, with the prior correction LAMOST, the difference in occurrence rates of sub-earth-sized planets(0.5--1 $R_{\oplus}$, P$\textless$30 days, and $\eta$ $>$ 0.1) around high-\textit{V} and low-\textit{V} stars is only 0.77 $\sigma$. This small difference may be attributed to the small number of selected planets in this method of correction. Because occurrence rate of sub-earth-sized planets around medium-\textit{V} stars is smaller than that around low-\textit{V} stars, which means the correlation between occurrence rates of sub-earth-sized planets and stellar relative velocities may not be a simply anti- or positive- correlation. We could not draw the conclusion that occurrence rates of sub-earth-sized planets decrease with decreasing stellar relative velocities. However, occurrence rates of sub-earth-sized planets around high-\textit{V} stars is higher  than both of planets around medium-\textit{V} and low-\textit{V} stars, no matter which way we use to correction. Thus, the conclusion is relatively robust.

\subsubsection{Correction of occurrence rate of planets with radius of 4--20 $R_{\oplus}$} \label{cor 4-20}

\begin{figure}
    \centering
    \includegraphics[width=0.95\linewidth]{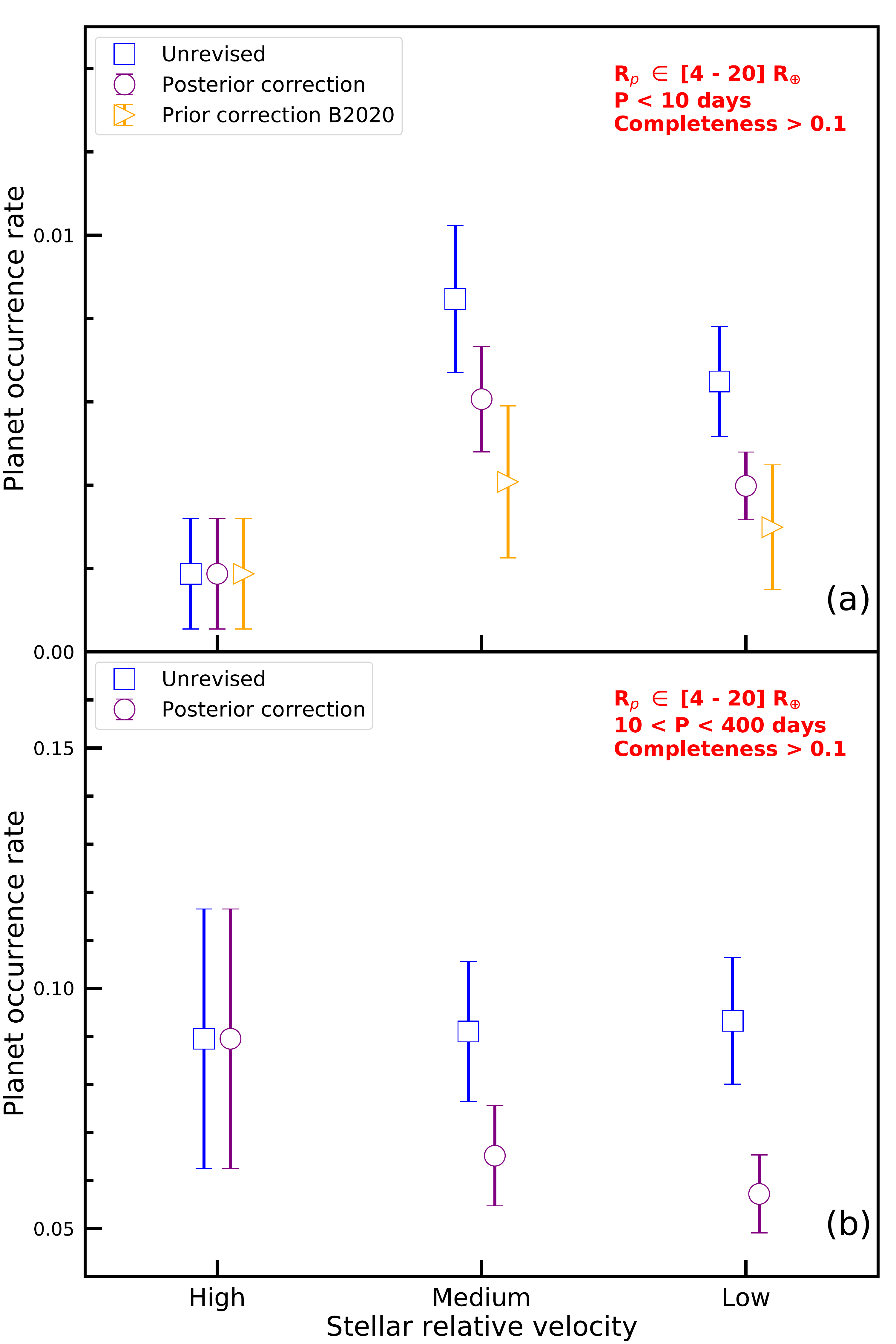}
    \caption{The occurrence rate of planets with a radius of 4--20 $R_{\oplus}$ around stars with different relative velocities. The planet occurrence rates are calculated in the orbital period of smaller than 400 days. Blue squares show the unrevised planet occurrence rate. Purple circles show the planet occurrence rate after Posterior correction. Orange triangles show the planet occurrence rate with Prior correction B2020.}
    \label{figure 3-7}
\end{figure}
For planets of $R_{\rm p}$ $\in$ 1--4 $R_{\oplus}$ and P $\textless$ 100 days, planet occurrence rate is anti-correlated with stellar relative velocity. While for planets of $R_{\rm p}$ $\in$ 0.5--1 $R_{\oplus}$ and P $\textless$ 30 days, high-\textit{V} stars have higher occurrence rate of these planets than low-\textit{V} stars. Here we also want to investigate the correlation between occurrence rate of planets of radius of 4--20 $R_{\oplus}$ and stellar relative velocity. For relatively small planets with radius of 0.5--4 $R_{\oplus}$, because the dependence of their occurrence rate on metallicity is much weaker than gas giants \cite{2019ApJ...873....8Z,2019ApJ...874...91W,2020arXiv200308431K}. Therefore, we change the empirical Equation (\ref{equation 3-4}) fitted from \cite{2019ApJ...873....8Z} to the empirical $f_{\rm occ}-[Fe/H]$ relations proposed by \cite{2010PASP..122..905J},

\begin{equation} \label{equation 3-7}
\footnotesize
    f(M_{\rm *},[Fe/H]) = 0.07 \pm 0.01 \times \left(\frac{M_{\rm *}}{M_{\rm \odot}}\right)^{1.0 \pm 0.3} \times 10 ^{1.2 \pm 0.2 [Fe/H]}.
\end{equation}
We use planets of the radius of 4--20 $R_{\oplus}$ because of the limited number of planets of 6 - 20 $R_{\oplus}$. Here, we assume that planets of the radius of 4--20 $R_{\oplus}$ share the similar $f_{\rm occ}-[Fe/H]$ relation with Jovian sized planets. Under this assumption, we also follow the methods in Appendix \ref{sub 3.3} to obtain the planet occurrence rate after correction. 

Several works have investigated the correlation between occurrence of Hot Jupiters and stellar velocities. For example, \cite{2019AJ....158..190H} show that hot jupiter(HJs) host stars have a smaller Galactic velocity dispersion than a similar population of stars without HJs, which implies that stars with a smaller Galactic velocity dispersion(similar to low-\textit{V} stars in my paper) may have a higher occurrence rate of HJs. Interestingly enough, \cite{2020Natur.586..528W} found that HJs prefer to exist around host stars in phase space overdensities.

Here we simply assume the planets with 4--20 $R_{\oplus}$ and P $\textless$ 10 days as HJs and the planets with 4--20 $R_{\oplus}$ and 10$\textless$P$\textless$400 days as warm Jupiters(WJs) or cold Jupiters(CJs). Fig \ref{figure 3-7} shows the correlation between the occurrence rate of Jupiter-sized planets and stellar relative velocity before and after the correction. Panel (a) shows the case of HJs and Panel (b) shows the case of WJs or CJs. 

Before the correction, the occurrence rates of HJs around medium-\textit{V} and low-\textit{V} stars are higher than that of HJs around high-\textit{V} stars with 2.99 $\sigma$ and 2.46 $\sigma$ respectively. Because the occurrence rate of Jupiter-sized planets have significantly positive correlation with metallicity, both the posterior correction and prior correction will reduce the occurrence rate of HJs around medium-\textit{V} and low-\textit{V} stars if we set high-\textit{V} stars as a control sample. After posterior correction, the occurrence rates of HJs around medium-\textit{V} and low-\textit{V} stars are higher than that of HJs around high-\textit{V} stars with 2.29 $\sigma$ and 1.36 $\sigma$ respectively. After prior correction B2020, the differences reduce to 0.98 $\sigma$ and 0.56 $\sigma$. In panel (b), we can see that before correction, occurrence rates of WJs or CJs are nearly the same with the changing stellar relative velocities. Yet after posterior correction, occurrence rate of WJs or CJs around high-\textit{V} stars is slightly higher than that of WJs or CJs around medium-\textit{V} and low-\textit{V} stars.

Although the occurrence rate of WJs and CJs is not significantly different, the enhanced occurrence rate of HJs around stars with lower relative velocites (similar to stars in phase space overdensities is consistent with previous works, i.e. \cite{2020Natur.586..528W}). Therefore, our results imply that clustering environment may play important role in the formation and evolution of HJs.

\subsection{statistical results of multiplicity and eccentricity}
\label{results of multiplicity and ecc}

Recently, \cite{2021arXiv210301974L} find that stars in high stellar phase space density environments (overdensities) have a factor of 1.6–2.0 excess in the number of single planet systems
compared to stars in low stellar phase space density environments (i.e.  field star). Their result suggests that stellar clustering may play an important role in shaping the properties of planetary systems, e.g. multiplicity, average eccentricity, and mutual inclination. Here, we want to investigate correlations between multiplicities, average eccentricities of planetary systems, and stellar relative velocities. We use two definitions to describe the multiplicity of planetary system, one is the fraction of multi-planets systems over the fraction of single planet systems, the other is the average number of planets per system. We will discuss the average number of planets per system, the fraction of multi-planets systems over the fraction of single planet systems, and the average eccentricity of planetary systems,respectively, in the following subsections.


\subsubsection{Average planet numbers per planetary system} \label{sub 5.1}
\begin{figure}
   \centering
    \includegraphics[width=0.95\linewidth]{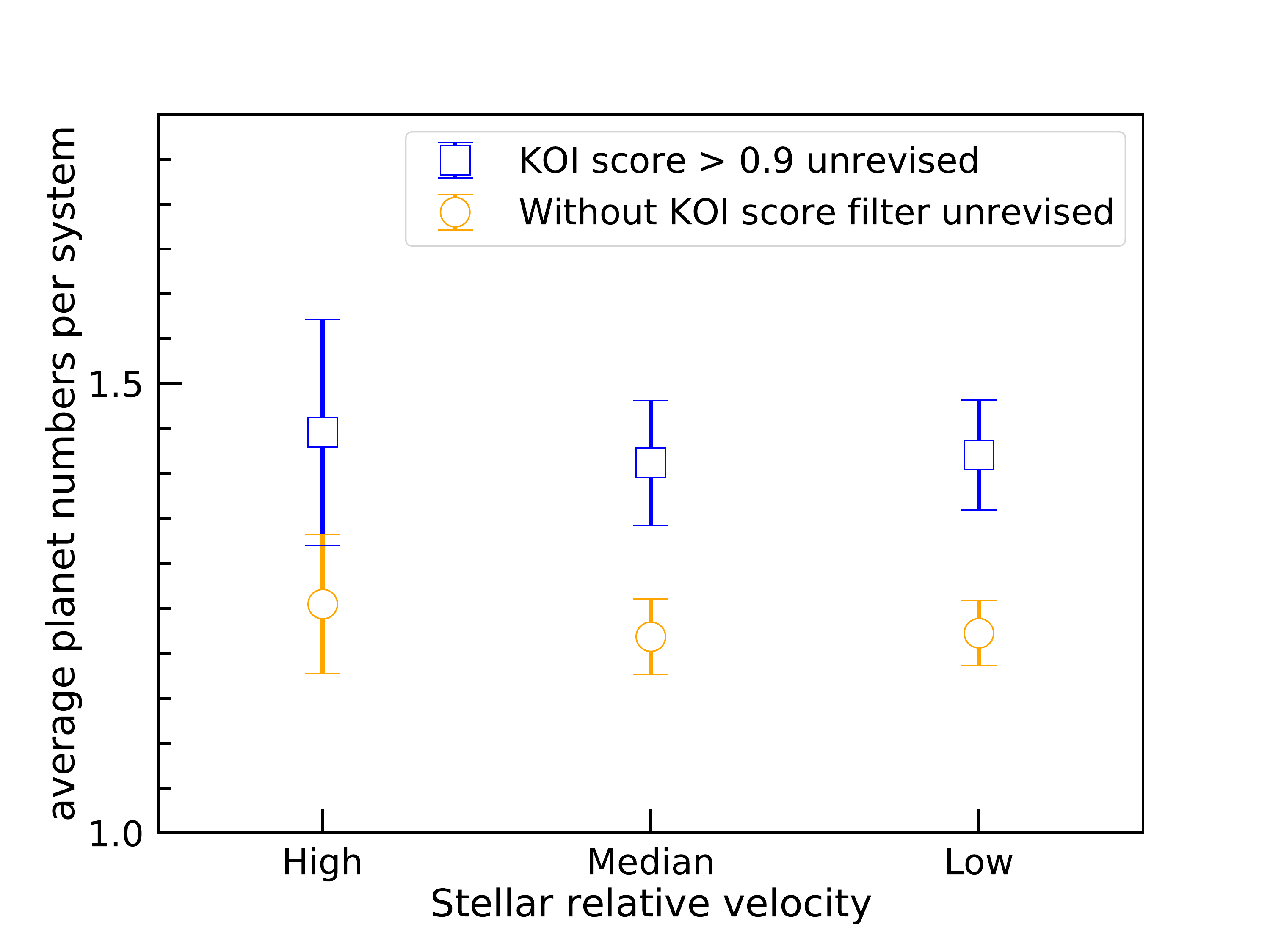}
    \caption{The average planet numbers per planetary system for planet hosts with different relative velocities. Blue symbols show the results of planets with koi score $\textgreater$ 0.9, while orange symbols show the results of planets without koi score filter. The error bar is calculated according to Poisson distribution.}
    \label{figure 5-1}
\end{figure}
The average planet numbers per planetary system is $\bar{N_{\rm p}}$,

\begin{equation} \label{equation 5-1}
    \bar{N_{\rm  p}} = \frac{\Sigma_{\rm i=1}^{n_{\rm  host}} N_{\rm i}}{n_{\rm  host}}
\end{equation}
where n$_{\rm  host}$ is the total number of planet hosts in the stellar samples, N$_{\rm i}$ is the number of transit planets around a given planet hosts. We use the Poisson distribution to calculate the error. In Fig \ref{figure 5-1}, considering the relatively large error of $\bar{N_{\rm  p}}$ of low-\textit{V} planet hosts, whether using the koi score filter or not, the average planet numbers per planetary system nearly unchanged with stellar relative velocity. It seems that the average planet numbers per planetary system have no significant correlation with stellar relative velocity. However, because planets with high mutual inclinations may not be detected by transit. Our calculated value of average planet numbers per planetary system is underestimated. In other words, "single" planet systems may not be single, multi-planet systems may have some other planets that haven't been detected due the limitation of observation time. Furthermore, if the stellar relative velocity is correlated with this underestimation of average planet numbers per planetary system, the statistical results may also be biased. 

Therefore, in order to minimize observational selection effects, in the next subsection, we will discuss another definition of the multiplicity of planetary systems, i.e. the fraction of multi-planets systems over the fraction of single planet systems, which is a relative value.


\subsubsection{The fraction of multi-planets systems over the fraction of single planet systems} \label{sub 5.2}
\begin{figure}
    \centering
    \includegraphics[width=0.95\linewidth]{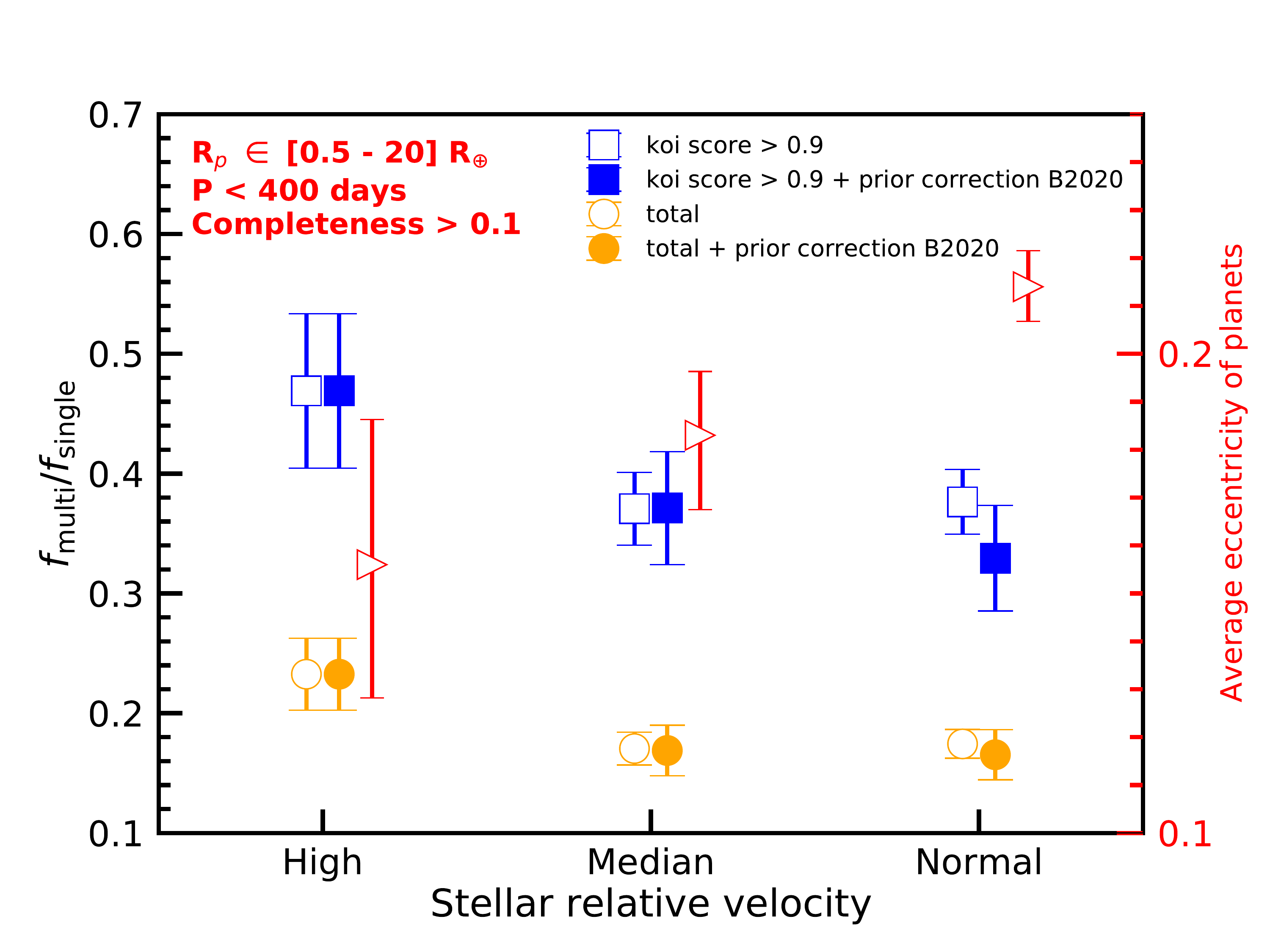}
    \caption{The correlations between multiplicities of planetary systems, average eccentricities of planets, and stellar relative velocities. Blue hollow squares show the uncorrected result of planets with koi score $\textgreater$ 0.9, while orange hollow squares show the uncorrected result of planets without koi score filter. The solid symbol shows the results of prior correction utilizing the metallicity of B2020. The red hollow triangles show the results of average eccentricities of planetary systems, i.e. eccentricity-velocity correlation. The error bar is calculated according to Poisson distribution. We select planets with the following criteria: (i), R$_{\rm p}$ $\in$ [0.5 - 20] $R_{\oplus}$; (ii), P $<$ 400 days; (iii) detection completeness $\eta$ $>$ 0.1.}
    \label{figure 5-2}
\end{figure}
Here we define multi-planet systems as stars that host more than one Kepler transit planet, and single-planet systems as stars that host only one Kepler transit planets. With the calculation of $f_{\rm  multi}/f_{\rm  single}$, i.e. a relative value which can avoid the influence of detection efficiency of different stellar samples. 

\begin{equation} \label{equation 5-1-1}
    \frac{f_{\rm  multi}}{f_{\rm  single}} = \frac{n_{\rm  multi} / n_{\rm  total}}{n_{\rm  single} / n_{\rm  total}}
\end{equation}
where $n_{\rm  multi}$, $n_{\rm  single}$ and $n_{\rm  total}$ are the number of multi-planet systems, single-planet systems and total planet systems in three stellar sub samples. 

Actually, some other planets with high mutual inclinations can not be discovered by transit. These "single" planet systems are not really single. Here, because we use a relative value to discuss the correlation between the multiplicity of a planetary system and stellar relative velocity, even if the transit method has some unknown preference on stellar relative velocity, this relative value can also minimize this secondary influence on the correlation. Thus we could draw a relatively robust conclusion. 

In figure \ref{figure 5-2}, with three selection criteria, i.e. (i), R$_{\rm p}$ $\in$ [0.5 - 20] $R_{\oplus}$; (ii), P $<$ 400 days; (iii) detection completeness $\eta$ $>$ 0.1, we show the correlation between $f_{\rm  multi}/f_{\rm  single}$ and stellar relative velocities both before and after the correction, i.e. Both hollow symbols(correction) and solid symbols(after correction) show that planetary systems around high-\textit{V} stars have a higher multiplicity than low-\textit{V} stars. The case with koi score filter shows a significantly higher $f_{\rm  multi}/f_{\rm  single}$ than the case without. Choosing the case of blue hollow squares and blue solid squares as an comparison, we can easily find that after correction the difference between multiplicities of planetary systems around high-\textit{V} stars and low-\textit{V} stars is even larger, i.e. 1.3 $\sigma$ correction and 1.8 $\sigma$ after correction. 

\begin{figure*}
    \centering
    \includegraphics[width=0.95\linewidth]{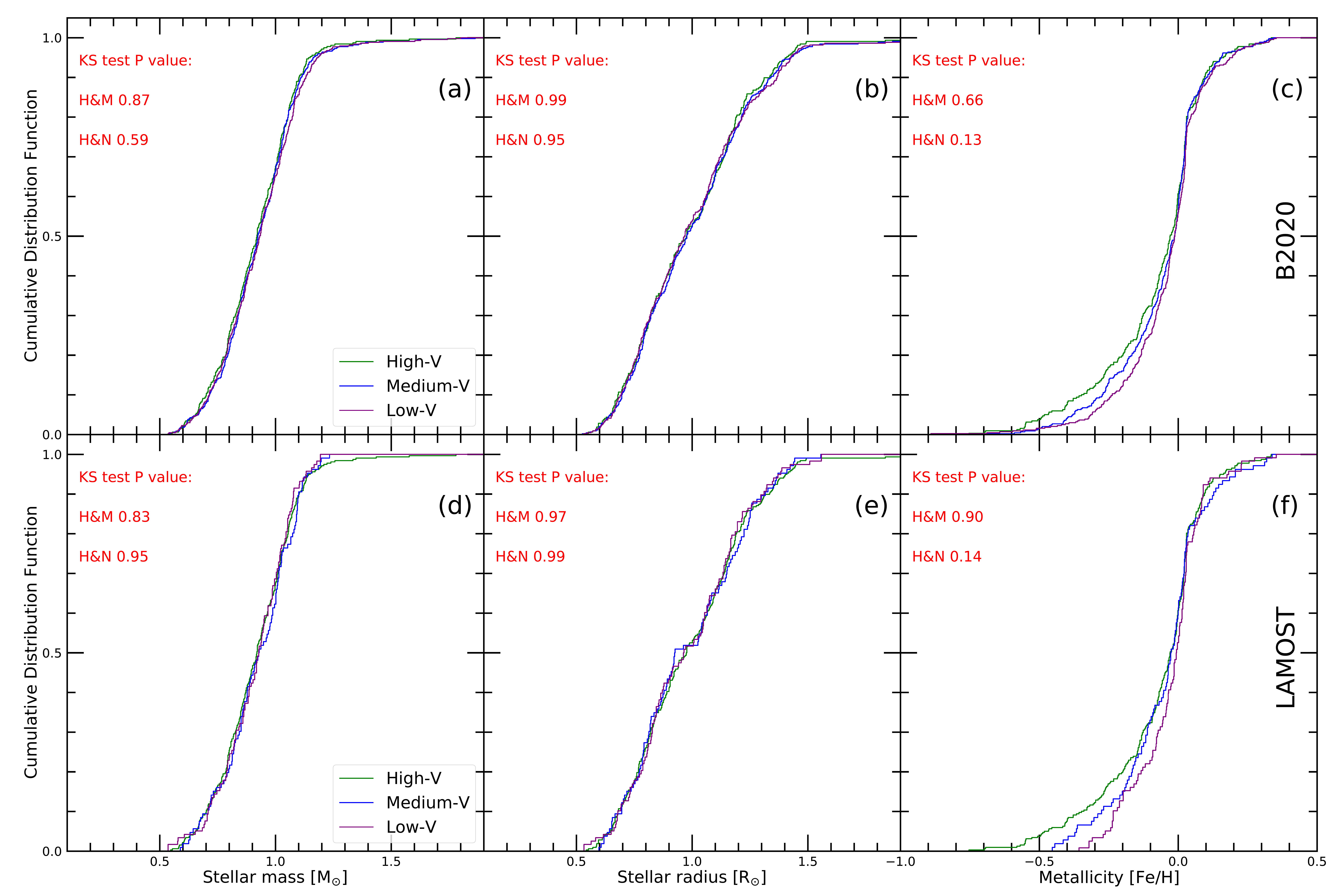}
    \caption{The cumulative distribution function(CDF) of stellar mass(panel (a)), radius(panel (b)) and metallicity of B2020(panel (c)), for high-\textit{V} , medium-\textit{V} and low-\textit{V} planet hosts respectively. We select medium-\textit{V} and low-\textit{V} planet hosts with the nearest stellar properties of every given high-\textit{V} planet host. Green, blue and purple lines show high-\textit{V} and those selected medium-\textit{V} and low-\textit{V} planets hosts respectively. Panels (d)(e)(f) are similar to Panels (a)(b)(c). The difference lies in the metallicity, i.e. panels (d)(e)(f) use planets hosts with metallicity of LAMOST DR4, while panels (a)(b)(c) use planets hosts with metallicity of B2020.}
    \label{figure 5-3}
\end{figure*}

The different stellar properties in different stellar sample may influence our results. In order to minimize such impact, here we also used the NearestNeighbors function in scikit-learn \citep{JMLR:v12:pedregosa11a} to choose the two medium-\textit{V} or low-\textit{V} planet hosts with nearest value of stellar mass, radius, and metallicity of B2020 for every selected high-\textit{V} planet, i.e. Prior correction B2020. Taking the case of no koi score filter as an example, after selecting the nearest medium-\textit{V} or low-\textit{V} planets host for every high-\textit{V} planets host, we calculate the $f_{\rm  multi}/f_{\rm  single}$, for medium-\textit{V} , $f_{\rm  multi}/f_{\rm  single}$ = 0.164$_{\rm -0.024}^{+0.024}$, for low-\textit{V} , $f_{\rm  multi}/f_{\rm  single}$ = 0.167$_{\rm -0.024}^{+0.024}$. These two values are nearly the same as values calculated with total medium-\textit{V} and low-\textit{V} planet hosts. In Fig \ref{figure 5-3}, we used K-S two sample test to compare distribution of stellar mass, radius and metallicity for high-\textit{V} and selected medium-\textit{V} and low-\textit{V} planet hosts, producing p-values of 0.87 or 0.59, 0.99 or 0.95 and 0.66 or 0.13 respectively. Therefore we confirm that the distributions are statistically similar. The solid symbols clearly show that high-\textit{V} planet hosts have higher $f_{\rm  multi}/f_{\rm  single}$ than low-\textit{V} planet hosts after the prior correction B2020. Here we do not list the results of prior correction LAMOST because of the limited number of selected planet hosts with the metallicity of LAMOST DR4. If don't take into account the large uncertainty due to the small number, after prior correction with the metallicity of LAMOST, the fractions of multi-planets systems over the fraction of single planet systems(koi score $\textgreater$ 0.9) are 0.5, 0.44, and 0.3 for high-\textit{V} , medium-\textit{V} and low-\textit{V} planet hosts respectively. The KS two-sample test also shows that for prior correlation LAMOST(Fig \ref{figure 5-3}), our selected medium-\textit{V} and low-\textit{V} planet hosts have statistically similar distributions with high-\textit{V} planet hosts.

\subsubsection{The average eccentricity of planets} \label{sub 5.3}
Here we use a robust general method to derive eccentricity distribution which is based on the statistics of transit duration \cite{2008ApJ...678.1407F} -- the time for transit planets to cross the stellar disks. Kepler's second law states that eccentric planets vary their velocity throughout their orbit. This results in a different duration for their transits relative to the circular orbit. Using equation (1) in \cite{2016PNAS..11311431X} and infer the eccentricity distribution from the statistics of $t_{dur}/t_{dur,0}$ ($t_{dur,0}$ is the transit duration for a circular orbit). Then, we could obtain the average eccentricity of planets around stars with different relative velocities.



The result of eccentricity is also shown in figure \ref{figure 5-2}. The average eccentricity of planets increases with decreasing stellar relative velocities, which show a significant anti-correlation with the trend of multiplicity. More specifically, planets around low-\textit{V} stars have larger average eccentricity than planets around high-\textit{V} stars with 2 $\sigma$ difference. The eccentricity dichotomy states that Kepler singles are on eccentric orbits with $\bar{e} \approx 0.3$, while the multiples are on nearly circular ($\bar{e} = 0.04_{\rm -0.04}^{+0.03}$) and coplanar ($\bar{i} = 1.4_{\rm -1.1}^{+0.8}$ degree) orbits \cite{2016PNAS..11311431X}. Therefore, our result is consistent with eccentricity dichotomy. 

Recent work \cite{2021arXiv210301974L} find single planetary systems prefer to exist around stars in overdensities with full \cite{2020Natur.586..528W} sample. We firstly find an anti-correlation between eccentricity and stellar relative velocities. Again, with the assumption that low-\textit{V} stars are similar to stars in overdensities and high-\textit{V} stars are similar to field stars, our finds also imply the significant influence of stellar clustering on the architecture of planetary systems. 

\section{Discussion}
\label{scenarios}
In this section, we mainly discuss the potential interpretations on these statistical results. 

Recently, several works point out that stellar clustering may play an important role in the formation and evolution of planets \citep{2020Natur.586..528W,2020ApJ...905L..18K,2021arXiv210301974L,2021ApJ...910L..19C}. Furthermore, \cite{2013ApJ...772..142L} found that most planetary systems ejected from open clusters maintains the original planetary architectures. Yet the planetary systems remain in clusters may be influenced by dynamical mechanisms significantly. Here we try to establish the correlation between our separations (i.e. high-\textit{V} , medium-\textit{V} , and low-\textit{V} stars) and those classifications (i.e stars in phase space overdensities and field stars) qualitatively. Stars in phase space overdensities are stars have similar positions and 3D velocities. In our sample selection, high-\textit{V} , medium-\textit{V} , and low-\textit{V} stars have similar positions, yet the different relative 2D velocities. Therefore, we could assume that the low-\textit{V} stars are similar to stars in phase space overdensities and the high-\textit{V} stars are similar to field stars. 

With such assumption, our results of Jupiter-sized planets are consistent with previous works. For example, \cite{2020Natur.586..528W} HJs are almost exclusively found in overdensities. We found a slightly enhancement of occurrence rate of HJs around low-\textit{V} stars. Additionally, we also find that high-\textit{V} stars have a slight lower occurrence rate of WJs or CJs than low-\textit{V} stars.   

There are two simple ways to explain our statistical results from the origin of Jupiter-sized planets. One is that high-\textit{V} stars have a higher in-situ formation rate of WJs or CJs while a lower in-situ formation rate of HJs. The other is forming ex situ(including disk migration and high-eccentricity migration), i.e. high-\textit{V} stars have a lower efficiency of forming from CJs to HJs than low-\textit{V} stars. Here we focus on the high-eccentricity migration. Dynamical perturbation can easily trigger the difference of eccentricity and multiplicity of planetary systems. Thus, high-eccentricity migration may be a more reasonable scenario to answer several our statistical results.    

In subsection \ref{high-e}, we will discuss several mechanisms that can trigger higher-eccentricity migration. In subsection \ref{test ecc} and \ref{test small planets}, we will use our results of small planets, multiplicity and eccentricity to test these mechanisms.


\subsection{High-eccentricity migration correlated to stellar relative velocities} \label{high-e}
According to the source of dynamical perturbation, we classify the high-eccentricity migration of HJs as three types: Stellar binary Kozai \citep{2016MNRAS.460.1086M}, Stellar flyby Kozai \citep{2017AJ....154..272H,2021arXiv210207898R}, and Planet-planet interactions(planet secular coplanar \citep{2015ApJ...805...75P}, Planet-planet Kozai \citep{2016ApJ...829..132P}, and Planet-planet scattering) \citep{2012ApJ...751..119B}. Because our difference of occurrence rate of HJs is correlated to stellar reative velocities, we prefer scenarios including external dynamical perturbations, i.e. Stellar binary Koizai and Stellar flyby Kozai. We also do not focus on the mechanism of planet secular coplanar, since it does not require the a proto-eccentric planet or an external source perturbation. For the mechanism such as planet-planet Kozai and planet-planet scattering, they may work after the previous external perturbation. E.g. stellar flyby could strongly excite the orbital elements of planets in the outermost orbit and the effect propagates to the entire planetary system through secular planet-planet interaction \cite{2017MNRAS.470.4337C}.  

\begin{figure}
    \centering
    \includegraphics[width=0.95\linewidth]{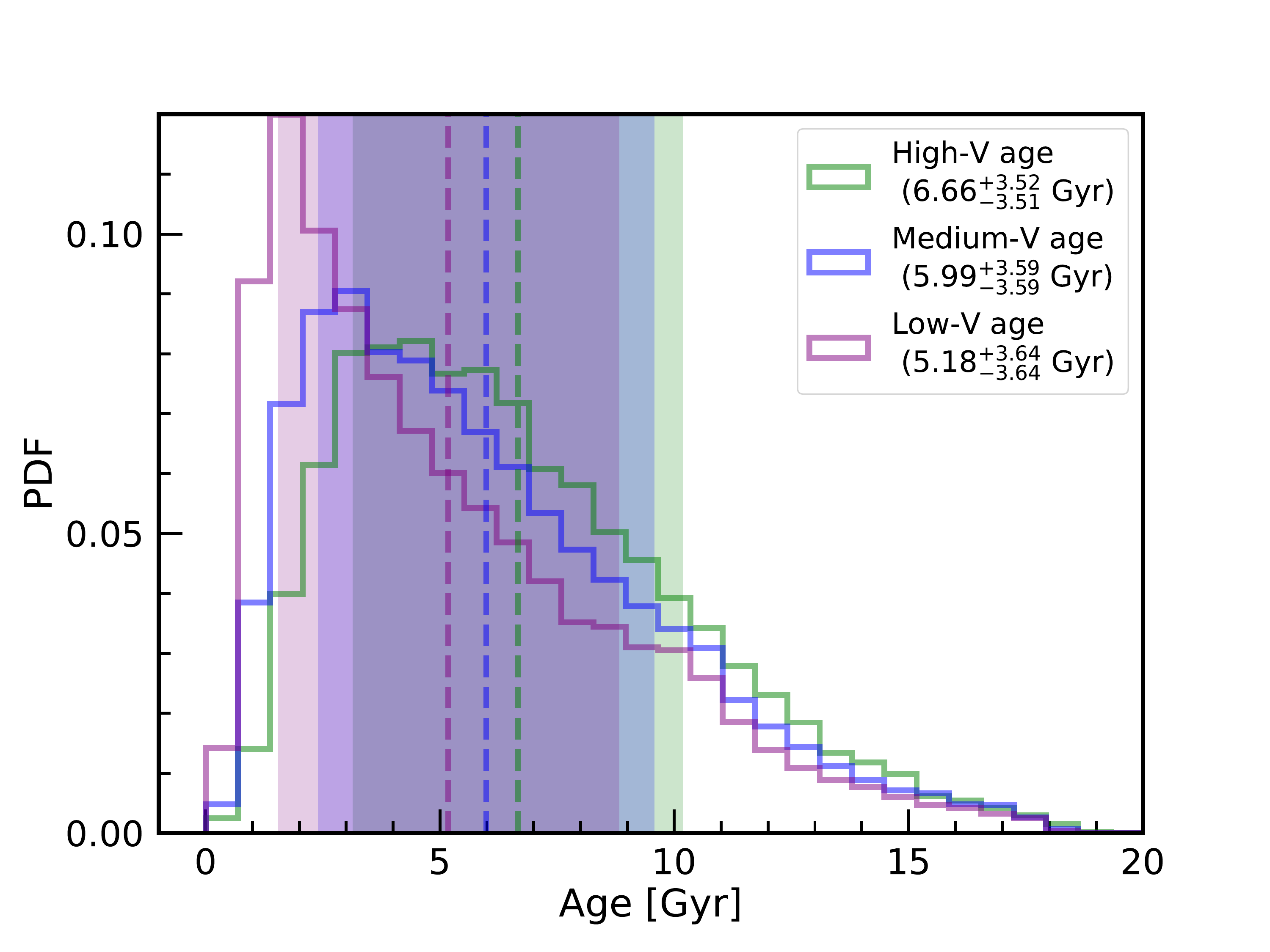}
    \caption{The distribution of stellar age listed in \cite{2020AJ....159..280B}. Green, blue and Purple shows high-\textit{V} , medium-\textit{V} and low-\textit{V} stars respectively. Solid lines are probability distribution functions. Dashed lines show the average age of stars. The colored regions show the standard deviation of stellar age.}
    \label{figure 6-1}
\end{figure}

\cite{2019AJ....158..190H} found that hot Jupiter host stars have a smaller Galactic velocity dispersion than a similar population of stars without hot Jupiters. According to the age-velocity dispersion relation (AVR) \citep{1946ApJ...104...12S,1977A&A....60..263W,2004A&A...418..989N,2018MNRAS.475.1093Y}, hot Jupiter host stars may be on average younger than field stars. For our stellar separation, we find that high-\textit{V} stars are older than low-\textit{V} stars on average(see figure \ref{figure 6-1}). Therefore, tidal migration of planets, which is correlated to age, may be another scenario to interpret our results of HJs. 

We prefer the dynamical perturbation instead of the age-dependent tidal dissipation. One reason is the large uncertainty of the age measurement. Furthermore, some clues are suggesting that low-\textit{V} stars may experience more environmental influence. \cite{2021A&A...647A..19T} shows that the heating rate of the Open clusters population is significantly lower for the vertical component compared to the field stars. And the age of clusters indicates some but weak age dependence of the known moving groups. Additionally, both simulation and observation suggest that a high fraction of comoving stars with small physical and 3D velocity separation are conatal \citet{2019ApJ...884L..42K}. These works imply that star groups can keep low relative velocities for a long timescale as several Gyrs. Thus the influence on relative velocity dispersion due to stellar age is considered less significant in this paper. Additionally, the dynamical perturbation can easily interpret our results of eccentricity and multiplicity(more details in section \ref{test ecc}).

\subsubsection{Binary induced high-eccentricity migration}
Recently, \cite{2020MNRAS.499.1212L} has shown that binaries are an important ingredient in understanding the importance of dynamical perturbation to planetary systems. Encounter rates for binaries may be larger than single stars. In typical open clusters, nearly 10 per cent of the Sun–Jupiter pairs acquire a stellar companion during scatterings. Such simulation results implies that the binary induced high-eccentricity migration is common in clusters.  

Although, there is no evidence that stars with low-\textit{V} have a denser birth environment e.g. dense clusters than high-\textit{V} stars, we can qualitatively give the  effective interaction rate for disruption under their current stellar environment. As described in \cite{2015MNRAS.448..344L}, the effective interaction rate for disruption is,

\begin{equation} \label{equation 9}
    \Gamma = n_{*}\langle \sigma \rangle \langle v \rangle, 
\end{equation}
where $n_{*}$ is the mean density of stars in the environment, $\langle v\rangle$ is the mean relative velocity between systems, and $\langle \sigma \rangle$ is the cross-section for the given mode of disruption. Because of the velocity dependence of the cross-sections(i.e.$\langle \sigma \rangle _{v} \equiv \langle \sigma v \rangle/\langle v \rangle$), high-\textit{V} stars with higher speed and lower effective interaction rate are less effected by passing flybys than low-\textit{V} stars. 

Similarly, according to the equation(11) in \cite{2020MNRAS.499.1212L}, the rate of Sun–Jupiter pairs acquiring a stellar companion during scatterings also depends on velocity. Therefore, high-\textit{V} stars with higher speed will result in the lower occurrence rate. I.e. high-\textit{V} stars could obtain companions with lower efficiency than low-\textit{V} stars and lower occurrence rate of HJs(induced by binary companion perturbation) consequently. Additionally, Jupiters may be directly ejected during such close encounters, although the fraction of these ejected Jupiters make up only 1\%. These ejection events may happen more frequently around low-\textit{V} stars due to velocity dependence, which will also support our statistical results of Jupiter-sized planets.


\subsubsection{Flyby induced high-eccentricity migration}
\cite{2020MNRAS.499.1212L} argued that close encounters for binaries may be larger than single stars, i.e. binaries may be the dominant scource of perturbation during the flyby events. Yet, single flyby events still occupy a significant fraction. \cite{2017AJ....154..272H} proposed that HJs could be driven by high-eccentricity migration in globular clusters. They found $\sim $ 2\% of giant planets that are converted to HJs, for the intermediate stellar density 10$^{4}$ pc$^{-3}$, i.e. flyby induced high-eccentricity migration \cite{2021arXiv210207898R}. Half of these giant planets may be ejected from the original systems. Similar to the analysis in the scenario of binary induced high-eccentricity migration, high-\textit{V} stars with higher relative velocities will have lower rate of, no matter the ejection of giant planets or HJs forming from CJs than low-\textit{V} stars. Thus, flyby induced high-eccentricity migration may be another channel to explain our statistical results of Jupiter-sized planets(with the assumption that stellar relative velocities correlate to stellar clustering).

\subsubsection{Planet-planet interaction}
Addition to dynamical perturbation from external sources, planet-planet interaction, e.g. planet–planet Kozai \citep{2016ApJ...829..132P} and planet-planet scattering \citep{2012ApJ...751..119B} can also contribute the formation of HJs. However, these two mechanisms have some preconditions. Planet-planet Kozai requires the initial mutual inclinations of the two giant planets, while planet-planet scattering requires three giant planets whose system usually goes unstable. Here we focus on the planet-planet interaction after the flyby events in clusters. For example, \cite{2017MNRAS.470.4337C} found that the orbital elements during flyby events are most strongly excited in the outermost orbit and the effect propagates to the entire planetary system through secular evolution. This so-called planet-planet interactions after flyby events or some other dynamical perturbation could be another way to HJs formation. Due to the additional conditions compared to the other two scenarios, we consider planet-planet interaction as a secondary influence. However, planet-planet interaction may play an important role in answering the statistical results of smaller planets, multiplicity, and eccentricity of planetary systems.

\subsection{Explaining the results of multiplicity and eccentricity}\label{test ecc}
The perturbation of external source can excite the mutual inclinations and eccentricities of planetary systems. \cite{2017MNRAS.470.4337C} found that clusters with higher stellar density can trigger the higher of average eccentricity of planetary systems with a larger eccentricity dispersion than clusters with lower stellar density. Additionally, they also find the eccentricity-multiplicity dichotomy similar to observation results \cite{2016PNAS..11311431X} and consistent with our statistical results.

If we assume the same velocity dispersion and the same mean relative velocity of stars in clusters with different stellar density, the rate of flyby events will be dominating by stellar density(see equation \ref{equation 9}). Therefore, higher rate of flyby events, the larger of mutual inclinations and average eccentricities of planetary systems. Although we could not link the stellar relative velocity of stars to the stellar density of their parental clusters directly, we can use the occurrence rate of flyby events to connect these two parameters. As is discussed above, high-\textit{V} stars with higher relative velocity, have the higher rate of flyby events. \cite{2020Natur.586..528W} stars may remain co-moving state after the disruption of clusters for Gyrs i.e. a relatively higher stellar density compared with other field stars. Therefore, our eccentricity and multiplicity results imply low-\textit{V} stars may be related to higher density of their parental clusters, or at least may have experienced more dynamical perturbation events. 

\subsection{Explaining the results of occurrence rate of small planets}\label{test small planets}
High-eccentricity migration may contribute to the close in planets(i.e. hot Super-earths and hot Neptunes)\cite{2016MNRAS.460.1086M}. If the high-eccentricity migration is the dominating channel of forming such close in planets(1--4$R_{\oplus}$, P $<$ 100 days), no matter flyby induced or binary induced, we would expect the similar results results HJs, i.e. high-\textit{V} stars have a lower occurrence rate of close-in super-Earths(P$\textless$30 days) and a potentially higher occurrence rate of cold super-Earths(P $>$ 400 days). Additionally, the scenario of planet-planet interaction after dynamical perturbation on outer small planets may be a possible channel to to explain our statistical results.  Eccentricity growth via scattering is limited to an epicyclic velocity corresponding to the escape velocity from the surface of the planet(e.g. \cite{2013ApJ...775...42I,2014ApJ...786..101P}),
\begin{equation} \label{equation scattering}
\begin{split}
    e_{\rm scatter} & \lesssim \frac{\sqrt{2GM_{\rm p}/R_{\rm p}}}{2\pi a /P} \\
    &= 0.2 \left(\frac{M_{\rm p}}{0.5 M_{\rm Jup}}\right)^{1/2}\left(\frac{2R_{\rm Jup}}{R_{p}}\right)^{1/2}\left(\frac{P}{3 day}\right)^{1/3}
\end{split}
\end{equation}
Once the eccentricity reaches this value, the cross section for collisions exceeds the cross section for scattering, and planets tend to merge rather than scatter during close encounters \citep{2018ARA&A..56..175D}. Given that sub-Earths with the smaller size than super-Earths, those cold sub-Earths are more likely merging than super-Earths during the planet-planet scattering. This mechanism may contribute to the decline of occurrence rate of sub-earths with decreasing stellar relative velocities. 

Additionally, since sub-earth may be excited more easily than larger planets, especially in single planetary systems. Therefore, qualitatively speaking, small planets may have the higher of mutual inclination. Therefore, some of single planetary systems may have a sub-earths with a relatively higher mutual inclination that can not be detected by transit. Therefore, such excitation of inclination may also contribute to the lower occurrence rate of sub-Earths around low-\textit{V} stars.

\section{conclusion} \label{conclusion}
Stars experience extreme dynamical interaction may influence the planet formation and evolution significantly. To explore the influence of such dynamical history on planet formation and evolution. We choose the relative velocity as a diagnose to show its correlation with the planet occurrence rate, multiplicity, and average eccentricity of planetary systems.

 We carefully select 74002 main-sequence single Kepler stars(FGK type) and 1910 reliable planets with deposition score larger than 0.9 in subsection \ref{sub 2.1}. Then, we calculate the two-dimensional relative velocity of these selected stars based on Gaia DR2, in subsection \ref{sub 2.2}. We divide the stars into three groups, i.e. high-\textit{V} stars, medium-\textit{V} stars, and low-\textit{V} stars due to different relative velocity. 
 There are some correlations between stellar properties and stellar relative velocity, i.e. high-\textit{V} stars with higher relative velocity, have smaller average stellar mass, lower average effective temperature, and lower average stellar metallicity. Considering the correlations between stellar properties and relative velocity (e.g. stellar effective temperature and stellar metallicity), and the influence of stellar properties on planet occurrence rate, we utilize two methods to correct these selection biases, i.e. prior correction and posterior correction. 
 
 After correlations, we calculate the occurrence rate of planets around high-\textit{V} stars, medium-\textit{V} stars, and low-\textit{V} stars and find some interesting correlations between stellar relative velocity and planet occurrence rate, as well as multiplicity and eccentricity in section \ref{results of focc}. The main statistical results are listed in the following, 


\begin{itemize}
    \item high-\textit{V} stars have a lower occurrence rate of super-earth and sub-Neptune-sized planets(1--4 $R_{\oplus}$, P$\textless$100 days, $\eta$ $>$ 0.1) than low-\textit{V} stars on average. 
    \item high-\textit{V} stars have a higher occurrence rate of sub-earth-sized planets(0.5--1 $R_{\oplus}$, P$\textless$30 days, $\eta$ $>$ 0.1) than low-\textit{V} stars on average.
    \item high-\textit{V} stars have a slightly lower occurrence rate of HJs (4--20 $R_{\oplus}$, P$\textless$10 days) than low-\textit{V} stars on average. While for WJs or CJs (4--20 $R_{\oplus}$, 10$<$P$<$400 days), high-\textit{V} stars have a slightly higher occurrence of them than low-\textit{V} stars on average. 
    \item The multiplicity of planetary systems increases with increasing stellar relative velocities, while the average eccentricity of planets shows an anti-correlation with stellar relative velocities, consistent with the eccentricity-multiplicity dichotomy. 
\end{itemize}


In section \ref{scenarios}, we discuss several scenarios to explain our statistical results. Considering the age-velocity relation, i.e. older stars tend to the lower of, the enhancement of HJs around stars with low-\textit{V} is consistent with previous studies. High-eccentricity migration may be a possible mechanism to explain the formation of HJs. Because such dynamical mechanism may be related to stellar clustering, no matter the binary induced \citep{2020MNRAS.499.1212L} or flyby induced \citep{2021arXiv210207898R}. Clustering evolution coupled to high-eccentricity migration may be a likely channel to explain the results of Jupiter-sized planets. 

Furthermore, external dynamical perturbation, e.g. close encounters in cluster could easily excite mutual inclinations and eccentricities which may account for with eccentricity-multiplicity dichotomy \citep{2017MNRAS.470.4337C}. Interestingly, our finding of multiplicity of planetary systems and average eccentricity of planets may be related to clustering environments(see also in \cite{2021arXiv210301974L} whose results are consistent with ours). 

Stellar clustering may also play important roles in the formation and evolution of smaller planets e.g. super-Earths and sub-Neptunes \cite{2020ApJ...905L..18K,2021ApJ...910L..19C,2021arXiv210301974L}. Similar to the channel of HJ's formation, the enhancement of super-Earths and sub-Neptunes around low-\textit{V} stars may be related to the high-eccentricity migration, at least some of close-in planets can be attributed to. The lower of occurrence rate sub-earth-sized planets around low-\textit{V} stars might be cause of higher of mutual inclination angles in which case transit methods could hardly discover. Yet, whether our findings about those planets with radius of 0.5--4 $R_{\oplus}$ correlated to stellar clustering needs more observational and theoretical works.    
In the future, with TESS and PLATO, more planets will be discovered. A larger sample of exoplanets will help to convince the correlation between occurrence rate and stellar relative velocities. More Planets in different clusters also benefit us to know the essential influences on planet formation and evolution due to cluster environments. 


 
Fortunately, with the data release from Gaia, more giant planets can be detected with a relatively larger semi-major axis compared with recent transit planets. These planets with longer periods can not only extend our knowledge of giant planet formation but also be able to valid our results, i.e. the correlation of giant planet occurrence rate and stellar relative velocity. 




\acknowledgments
We thank the anonymous referee for helpful comments and feedback. We also thank Dr. Andrew Winter and Ji-Wei Xie for helpful comments and suggestions. This work is supported by the National Natural Science Foundation of China (Grant No. 11973028, 11933001, 11803012, 11673011), and the National Key R\&D Program of China (2019YFA0706601). The technology of Space Telescope Detecting Exoplanet and Life supported by the National Defense Science and Engineering Bureau civil spaceflight advanced research project D030201 also supports this work. This work made use of the stellar properties catalog of LAMOST DR4. Guoshoujing Telescope (the Large Sky Area Multi-Object Fiber Spectroscopic Telescope LAMOST) is a National Major Scientific Project built by the Chinese Academy of Sciences. Funding for the project has been provided by the National Development and Reform Commission. LAMOST is operated and managed by the National Astronomical Observatories, Chinese Academy of Sciences. This research made use of the cross-match service provided by CDS, Strasbourg. This research has made use of the NASA Exoplanet Archive, which is operated by the California Institute of Technology, under contract with the National Aeronautics and Space Administration under the Exoplanet Exploration Program.

%

\software{Astropy \citep{2013A&A...558A..33A},
          Matplotlib \citep{Hunter:2007},
          Scipy \cite{mckinney-proc-scipy-2010},
          Pandas \cite{reback2020pandas},
          Scikit-learn \citep{JMLR:v12:pedregosa11a}.
          }

\clearpage



\appendix
\renewcommand\thefigure{\Alph{section}\arabic{figure}}
\renewcommand\theequation{\Alph{section}\arabic{equation}}
\section{The difference between metallicity of B2020 and metallicity of LAMOST DR4} \label{app A}

\setcounter{figure}{0}
\begin{figure*}
    \centering
    \includegraphics[width=1\linewidth]{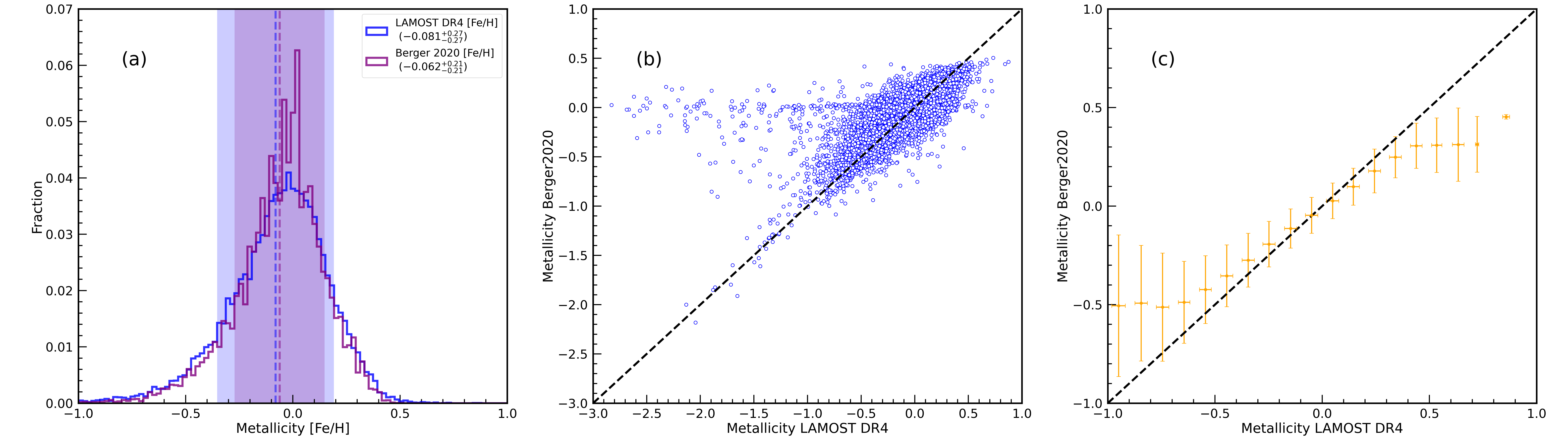}
    \caption{Comparison between stellar metallicity listed in LAMOST DR4 and B2020. Panel (a) shows the distribution of metallicity listed LAMOST RD4 and B2020. The blue and purple colors show the data of LAMOST DR4 and B2020 respectively. The solid line is the distribution of metallicity. The dashed line is the average value of metallicity. The colored region shows the standard deviation of metallicity. Panel (b) is a scatter diagram of stellar metallicity from LAMOST DR4 and metallicity from B2020. X-axis is stellar metallicity from LAMOST DR4. Y-axis is metallicity from B2020. Panel (c) shows the difference between stellar metallicity from LAMOST DR4 and metallicity from B2020. We bin the stellar metallicity from LAMOST DR4 with 0.1 indexes (the average standard error of metallicity LAMOST DR4). The error bar is the standard deviation of B2020 metallicity.}
    \label{figure 2-5}
\end{figure*}
In Fig \ref{figure 2-5}, panel (a) shows the distribution of stellar metallicity from LAMOST DR4(blue) and B2020(purple). The average stellar metallicity of LAMOST DR4(blue) is $-0.078_{\rm -0.28}^{+0.28}$, which is comparable with that of B2020  $-0.061_{\rm -0.21}^{+0.21}$. However, the stellar metallicity of B2020 has a higher fraction of around zero. It is more clear in panel (b) which shows the scatter diagram, where the X-axis shows metallicity([Fe/H]) of LAMOST DR4 and the Y-axis shows metallicity of B2020. Most of the values are distributed around the line(y=x), however, there is an abnormal branch around the [Fe/H]$_{\rm B}$=0, i.e. the metallicity of B2020. \cite{2020AJ....159..280B} mentioned that the metallicity ([Fe/H]$\textless$-0.5 and [Fe/H]$\textgreater$-0.5) have larger uncertainty. Consequently, this large uncertainty of input metallicity and uncertainty of the model will result in the different distribution of stellar metallicity (e.g. LAMOST DR4 and B2020) in some specific sub-sample, although their entire distribution of metallicity is nearly the same. In panel (c), we bin the LAMOST DR4 metallicity within 0.1 [Fe/H] index which is close to the standard error of LAMOST DR4 metallicity. The error bar is the standard deviation of metallicity from two catalogs in the bin. The dashed line fits well for metallicity $\in$ [-0.5, 0.5], which means the stellar metallicity in this range are consistent in both catalogs. While for metallicity outside the range of [-0.5, 0.5], LAMOST DR4 and B2020 shows a significant difference. 

For stars that do not have spectroscopic metallicity constraints ($\sim$120,000), \cite{2020AJ....159..280B} used a prior assumption that those stars have solar metallicity with a standard deviation of $\sim$0.20 index. Thus for those stars without constraints, the metallicity of B2020 will concentrate near zero. However, for those stars having LAMOST metallicity constraints, metallicity from B2020 should be similar to that from LAMOST DR4. LAMOST DR4 provides several catalogs of stellar metallicity from different pipelines. In our paper, we use the revised metallicities from \cite{2017MNRAS.467.1890X}, which may be different from the metallicity of B2020. Different values of metallicity may lead to different correlations, therefore, in section \ref{results of focc}, we will both discuss the influence of metallicity from different tables, on correlations between planet occurrence rate and stellar relative velocity. 
\section{Planet occurrence rate} \label{sub 2.3}
\setcounter{equation}{0}

The planet occurrence rate means the average number of planets per star. We follow \cite{2015ApJ...798..112M} to calculate the planet occurrence rate. Here we take into account the actual observation time of every selected Kepler star and calculate the signal-noise ratio and detection efficiency of planets with long orbital periods. Apart from that, we derive the formula for occurrence rate of planets around stars with stellar relative velocities. 

For a planet, its detection efficiency is different if it orbits around different stars. Before calculating the planet occurrence rate, we should first calculate the modeling signal-noise ratio of a given planet around different selected Kepler stars. The stellar noise presents in a light curve, the
so-called Combined Differential Photometric Precision (CDPP, \cite{2012PASP..124.1279C}) is time-varying. Here we use the robCDPP \cite{2015ApJ...809....8B}. Because non-robust rms CDPP (rmsCDPP) statistic typically reported by the Q1-Q16 pipeline data products can be biased. In some works, they use Poisson distribution to fit the noise of stars, while we follow the way of \cite{2015ApJ...798..112M} to assume the noise of stars follows a decaying power law. We use the data of 3,6 and 12 hours of robCDPP to fit the function between stellar noise $\sigma_{\rm *}$ and transition duration timescale $t$:

\begin{equation} \label{equation 2-5}
    \sigma_{\rm *} = \sigma_{\rm  LC} \left( \frac{t}{t_{\rm  LC}}\right)^{cdpp_{\rm  index}}, 
\end{equation}
where $\sigma_{\rm  LC}$ is the normalized stellar noise and $t_{\rm  LC}$(1765.5s) is the noise in the long cadence mode. 
For a planets with given orbital periods and planet radius, whether it can be detected is determined by the signal noise ratio(SNR). Noise is the noise of stars $\sigma_{\rm *}$, the signal can be simply considered as the transit depth $\delta$,

\begin{equation} \label{equation 2-6}
    \delta=\left(R_{\rm p}/R_{\rm *} \right)^{2},
\end{equation}
where R$_{\rm  \rm p}$ is the planet radius, R$_{\rm *}$ is the stellar radius. Because stellar noise is related with transit duration $t_{\rm  dur}$ and signal is approximately proportional to the square root of transit times $n$,

\begin{equation} \label{equation 2-7}
    n = \frac{t_{\rm obv}}{P},
\end{equation}
where $t_{\rm obv}$ is the observation time of a given Kepler star, and P is the orbital period of a planet. $t_{\rm obv}$ is written as, 

\begin{equation} \label{equation 2-8}
    t_{\rm obv} = duty cycle * data span.
\end{equation}
Signal noise ratio can be written as,

\begin{equation} \label{equation 2-9}
    SNR = \frac{\delta n^{0.5}}{\sigma \left( t_{\rm  dur}\right)}.
\end{equation}
Planet transit duration $t_{\rm  dur}$ is written as,

\begin{equation} \label{equation 2-10}
    t_{\rm  dur} = \frac{P R_{\rm *} \sqrt{1-e^{2}}}{\pi a},
\end{equation}
where $R_{\rm *}$ is the stellar radius and $e$ is planet orbital eccentricity. Semi-major axis $a$ can be written as, 

\begin{equation} \label{equation 2-11}
    a = \sqrt[3]{\frac{G M_{\rm *} P^{2}}{4 \pi^{2}}}.
\end{equation}

Using this method, we can correct the systematic increase of detection efficiency caused by planets with small orbital periods. Here we don't take into account the impact parameter $b$. The calculation of transit duration needs the orbital eccentricity. Unfortunately, there are few planetary systems with eccentricity. However, because the eccentricity of the most planet is less than 0.3, the whole difference in transit duration caused by eccentricity is not larger than 5\%. Thus we simply fix the average eccentricity of Kepler planetary systems to 0.1, which is in the range of eccentricity given by \cite{2011ApJS..197....1M} (0.1 - 0.25).

In Kepler pipeline, confirming a potential planet candidate needs to observe three transits. Here we follow the formula in \citep{2013ApJ...763...12B}, this efficiency increases with the transit times linearly. Such efficiency determined by the transit numbers $f_{\rm  n}$ can be written as,

\begin{equation} \label{equation 2-12}
    \begin{split}
    t_{\rm  obs} \leq 2P &: f_{\rm  n}  = 0, \\
    2P < t_{\rm  obs} < 3P &: f_{\rm  n} = \left(t_{\rm  obs}/P-2\right), \\
    t_{\rm  obs} \geq 3 &: f_{\rm  n} = 1. 
    \end{split}
\end{equation}

Kepler pipeline defines a transit signal when SNR is larger than 7.1. Although these selection criteria can exclude many false positive signals, yet it may also exclude some potential real signal with low signal noise ratio because of the limit of observation time. Here we follow the way of \cite{2015ApJ...798..112M}, in which the detection efficiency $f_{\rm  eff}$ is dependent on the SNR and is assumed as linear where SNR is in the range of 6 to 12:   

\begin{equation}  \label{equation 2-13}
    \begin{split}
        \rm {SNR} \leq 6 &: f_{\rm  eff} = 0,\\
        6 < \rm {SNR} \leq 12 &: f_{\rm  eff} = \frac{ {SNR} -6}{6}, \\
        \rm {SNR} > 12 &: f_{\rm  eff} =1. 
    \end{split}
\end{equation}

Then, we can calculate number of Kepler stars around which a given planet can be detected, i.e. stellar number $N_{\rm *}$, 

\begin{equation} \label{equation 2-14}
    N_{\rm *}\left(q, R_{\rm  p},P\right)=\Sigma_{\rm i=0}^{N_{\rm *}}\left(f_{\rm  eff,i} \cdot f_{\rm  n,i}\right),
\end{equation} 
Here we round $N_{\rm *}$ to integer. $N_{\rm *}$ is the function of $q$, R$_{\rm  p}$, planet radius, and orbital period $P$.

The detection completeness $\eta=N_{\rm *}\left(q, R_{\rm  p},P\right)/N_{\rm star}\left(q\right)$, where $N_{\rm star}$ is the total number of selected Kepler main-sequence single stars with specific range of relative stellar velocity(i.e. with different q). The detection completeness and the planets are one-to-one relation. The detection completeness of a given planet is the fraction of stars around which we can detect the planet.

Besides the efficiency related with transit numbers and SNR, when we calculate the planet occurrence, transit probability caused by the geographic structure of a planetary system should be taken into account inevitably. The formula of the transit probability is, 

\begin{equation} \label{equation 2-15}
	f_{\rm  geo} = \frac{R_{\rm  p}+R_{\rm *}}{a\left( 1- e^
	{2}\right)}
\end{equation}
where $\left(1-e^{2}\right)$ is a revising impact caused by ellipse orbital \cite{2008ApJ...679.1566B}.   

For a planet with given planet radius and orbital period, its occurrence rate can be written as,

\begin{equation} \label{equation 2-16}
	f_{\rm  occ}\left(q,R_{\rm p},P\right) = \frac{1}{f_{\rm  geo}N_{\rm *}\left(q,R_{\rm p},P\right)}.
\end{equation}
For planets with given range of planet radius and orbital period, we add the calculate occurrence rate $f_{\rm  occ}\left(q,R_{\rm  p},P\right)$ cumulatively. The dominant source of error is Poisson error, as opposed to measurement errors. Therefore we estimate the confidence interval the usual $1/\sqrt{N_{\rm exp}}$ approach(where $N_{\rm exp}$ is the number of planets in a given range of planet radius and orbital period).

\section{Correction of occurrence rate of planets due to different stellar properties} \label{sub 3.3}
\setcounter{equation}{0}
We will introduce two simple ways to correct the influence of different stellar properties on the planet occurrence rate In this subsection. One way is the prior correction i.e. we minimize the influence of stellar properties before the calculation of occurrence rate, while the other is posterior correction i.e. we correct the influence of stellar properties utilizing the empirical relations after the calculation of occurrence rate.

\subsection{Prior correction} \label{subsub 3.3.1}
For prior correction, we select the stars with similar stellar properties of high-\textit{V} stars in both medium-\textit{V} stars and low-\textit{V} stars to minimize the influence of stellar properties. Since we select stars with a similar distribution of stellar properties such as stellar radius, stellar mass, and metallicity, the difference of other parameters such as planet occurrence rate can be probably attributed to the difference of stellar relative velocity. Here we use the NearestNeighbors function in scikit-learn \citep{JMLR:v12:pedregosa11a} to choose the two nearest medium-\textit{V} or low-\textit{V} stars having similar stellar mass, radius, and metallicity compared with every high-\textit{V} stars. After stellar selection, we use the methods described in Appendix \ref{sub 2.3} to calculate the planet occurrence rate. As a consequence, we obtain the planet occurrence rate after prior correction.

\subsection{Posterior correction} \label{subsub 3.3.2}
For posterior correction, we minimize the influence of stellar effective temperature and stellar metallicity through empirical relations. The correlations between planet occurrence and stellar effective temperature and stellar metallicity have been studied broadly. Here we choose the influence of stellar effective temperature instead of the stellar mass because that the uncertainty of stellar effective temperature(\%3, or 112 K) is lower than that of stellar mass(\%7). 

We assume the planet occurrence rate as a function of planet radius $R_{\rm p}$, orbital period $P$, stellar effective temperature $T_{\rm eff}$ and stellar metallicity $[Fe/H]$, i.e. $f_{\rm occ}(R_{\rm p}\,,P\,,T_{\rm eff}\,,[Fe/H])$. Because we focus on the occurrence rate of planets in specific radius range. Thus after integration, we can rewrite the f$_{\rm occ}$ as $f_{\rm occ}(P\,,T_{\rm eff}\,,[Fe/H])$. If we assume that orbital period $P$, stellar effective temperature $T_{\rm eff}$ and stellar metallicity $[Fe/H]$ are three independent variables. f$_{\rm occ}$ can be written as,

\begin{equation} \label{equation 3-1}
        f_{\rm occ}(P,T_{\rm eff},[Fe/H]) = f_{\rm 1}(P) f_{\rm 2}(T_{\rm eff})
        f_{\rm 3}([Fe/H]),
\end{equation}
where f$_{\rm 1}$(P) is in a form of broken power law as is shown in many other previous studies \citep{2015ApJ...799..180S,2018AJ....156...24M,2020ApJ...891...12N}. 

\begin{equation} \label{equation 3-2}
        f_{\rm 1}(P) = c_{\rm 1}
        \begin{cases}
        (P/P_{\rm 0})^{a} \,, & P\,\leqslant \,P_{\rm 0}, \\
        (P/P_{\rm 0})^{b} \,, & P\,> \,P_{\rm 0},
        \end{cases}
\end{equation}
where P$_{\rm 0}$ is the broken point where the index of the power law is different, a and b are the power law index and c$_{\rm 1}$ is constant. However, we do not focus on the function of planet occurrence rate with planet orbital periods. We are interested in $f_{\rm 2}(T_{\rm eff})$ and $f_{\rm 3}([Fe/H])$. 

For $f_{\rm 2}(T_{\rm eff})$, here we use the empirical relation between planet occurrence rate and effective temperature proposed by \cite{2020AJ....159..164Y},

\begin{equation} \label{equation 3-3}
    f_{\rm 2}\left(T_{\rm eff}\right) = c_{\rm 2}\left(0.30 + \frac{0.43}{1+exp\left(\frac{T_{\rm eff}-6061}{161}\right)}\right).
\end{equation}
Since this formula is about average planet multiplicity $\bar{N}_{\rm p}$. While actually in our definition of planet occurrence, i.e, t, $f_{\rm occ} = F_{\rm Kep} \bar{N}_{\rm p}$, where $F_{\rm Kep}$ is the fraction of planetary systems in the Kepler stars. Because $F_{\rm Kep}$ and $\bar{N}_{\rm p}$ have the similar correlation with $T_{\rm eff}$ as is shown in \cite{2020AJ....159..164Y}, we simply use the formula of $\bar{N}_{\rm p}$(with a normalized efficiency $c_{\rm 2}$) which is also consistent with the relation between $f_{\rm occ}$ and $T_{\rm eff}$(see figure 9 in \cite{2020AJ....159..164Y}. The majority of the planet detected by Kepler is in the range of 0.5--4 $R_{\oplus}$. Therefore this Equation (\ref{equation 3-3}) is reasonable. Although different studies will derive different empirical $f_{\rm occ}-T_{\rm eff}$ relations, the entire correlations between planet occurrence rate and stellar effective temperature are similar. Utilizing different formula can only influence our results slightly. Therefore, we use the equation (\ref{equation 3-5}) typically. 

For $f_{\rm 3}([Fe/H])$, we derive the function from a recent study on the correlation between planet occurrence rate and stellar metallicity \cite{2019ApJ...873....8Z}. \cite{2019ApJ...873....8Z} found that occurrence rates of Kepler-like planets around solar-like stars have a slightly positive correlation with metallicity. We use the data in figure 3 of \cite{2019ApJ...873....8Z}. We simply fit the data with a function as follows, 

\begin{equation} \label{equation 3-4}
    f_{\rm 3}\left([Fe/H]\right) = c_{\rm 3} [Fe/H]^{\alpha} + c_{\rm 4},
\end{equation}
where $\alpha$ is the power law index and c$_{\rm 3}$ and c$_{\rm 4}$ are constant values. In our following calculation, we fix $\alpha$ to a typical value, $\alpha = 1$. In other words, it's a linear relation. Noting that the data in figure 3 of \cite{2019ApJ...873....8Z} does not exclude the influence of multi-planet systems or giant planets, thus it shows the upper limit of the positive correlation between planet occurrence rate and metallicity.

We take into consideration of distribution of stellar effective temperature and stellar metallicity, and function of planet occurrence rate in order to revise the influence of stellar effective temperature and stellar metallicity. Thus the Equation (\ref{equation 3-1}) can be rewritten as, 

\begin{equation} \label{equation 3-5}
    \begin{split}
    f_{\rm occ}&= \iint f_{\rm 1}(P)f_{\rm 2}\left(T_{\rm  eff}\right)PDF\left(T_{\rm  eff}\right)f_{\rm 3}\left([Fe/H] \right)
    PDF\left([Fe/H]\right) dT_{\rm eff}d[Fe/H]  \\ 
    &= C_{\rm [Fe/H]} C_{\rm T_{\rm eff}} f_{\rm 1}(P),
    \end{split}
\end{equation}
where PDF$(T_{\rm eff})$ is the probability distribution function of stellar effective temperature, and PDF([Fe/H]) is the probability distribution function of stellar metallicity. 

Here if we want to get the integrated planet occurrence rate after correction, we only need to calculate the efficiency $C_{\rm [Fe/H]}$ and $C_{\rm T_{\rm eff}}$. For example, if we know the original stellar sample with PDF$_{\rm 0}(T_{\rm eff})$ and PDF$_{\rm 0}([Fe/H])$ and the planet occurrence rate $f_{\rm occ,0}$, and we want to get the occurrence rate of planets around stars with PDF$_{\rm 1}(T_{\rm eff})$ and PDF$_{\rm 1}([Fe/H])$, i.e. $f_{\rm occ,1}$, 

\begin{equation} \label{equation 3-6}
    f_{\rm occ,1} = C_{\rm 0 \Rightarrow 1} f_{\rm occ,0}, 
\end{equation}
where $C_{\rm 0 \Rightarrow 1}=\frac{C_{\rm T_{\rm eff},1}C_{\rm [Fe/H],1}}{C_{\rm T_{\rm eff},0}C_{\rm [Fe/H],0}}$ is the correction efficiency.  $C_{\rm T_{\rm eff},0}$, $C_{\rm [Fe/H],0}$, $C_{\rm T_{\rm eff},1}$ and $C_{\rm [Fe/H],1}$ are efficiencies related with distribution of stellar effective temperature and metallicity before and after the correction respectively.

\bibliography{my_paper}{}
\bibliographystyle{aasjournal}


\end{document}